\documentclass[12pt, a4paper]{article}
 \usepackage[font=small,format=plain,labelfont=bf,up,textfont=normal,up,justification=justified,singlelinecheck=false]{caption}
\usepackage{subcaption}

\usepackage[a4paper, left=2cm,right=2cm]{geometry}
\usepackage[colorlinks=true,linkcolor=black,citecolor=teal,urlcolor=MidnightBlue,filecolor=black]{hyperref}
\usepackage{amsfonts}
\usepackage{amsmath,amssymb}
\usepackage{setspace}
\usepackage[dvipsnames]{xcolor}
\definecolor{SchoolColor}{rgb}{0.6471, 0.1098, 0.1882} % Crimson

\usepackage[utf8,applemac]{inputenc}
\usepackage{tensor,braket}
\usepackage{cite}
\usepackage{tikz}
\usepackage{graphicx}% Include figure files
\usepackage{bm} % For bold math symbols

\usepackage[utf8,applemac]{inputenc}
\usepackage{tensor}
\usepackage{cite}
\usepackage{tikz}
\usepackage{graphicx}
\graphicspath{{figure/}}
\usepackage{array}
\usepackage{booktabs} %调整表格线与上下内容的间隔
\usepackage{multirow}

\usepackage{dcolumn}% Align table columns on decimal point
\usepackage{bm}% bold math

\usepackage{verbatim}

\usepackage{textcomp} % \textless, \textgreater, \textbrokenbar macros
\usepackage{graphicx} % \resizebox macro

\numberwithin{equation}{section}
\newcommand{\bea}{\begin{eqnarray}}
\newcommand{\eea}{\end{eqnarray}}
\newcommand{\be}{\begin{equation}}
\newcommand{\ee}{\end{equation}}
\def\nn{\nonumber}

\newcommand{\beqs}{\begin{eqnarray}}
\newcommand{\eeqs}{\end{eqnarray}}
\numberwithin{equation}{section}

\newcommand{\Rmnum}[1]{\uppercase\expandafter{\romannumeral #1\relax}}
\newcommand{\jl}{\color{green}}

\begin{document}
\begin{titlepage}

\begin{flushright}\vspace{-3cm}
{\small
%{\tt arXiv:yymm.nnnn} \\
\today }\end{flushright}
\vspace{0.5cm}
\begin{center}
	{{ \LARGE{\bf{Symmetry group at future null infinity \Rmnum{1} :
	
	Scalar theory}}}}\vspace{5mm}%\vspace{8pt}\\in higher dimensional CFT}}}} \vspace{5mm}

	\centerline{\large{\bf Wen-Bin  Liu\footnote{liuwenbin0036@hust.edu.cn}, Jiang Long\footnote{
				longjiang@hust.edu.cn}}}
	\vspace{2mm}
	\normalsize
	\bigskip\medskip
%	\textit{Asia Pacific Center for Theoretical Physics,\\ Pohang 37673, Korea}\\\
%\vspace{2mm}

	\textit{School of Physics, Huazhong University of Science and Technology, \\ Luoyu Road 1037, Wuhan, Hubei 430074, China
	}
	%\vfil
	%\pacs{04.70.Dy}
	
	\vspace{25mm}
	
  \begin{abstract}
		%{We reduce the massless scalar field theory in Minkowski spacetime to future null infinity. We compute the Poincar\'e flux operators at future null infinity and construct the supertranslation and superrotation generators. The generators are shown to form a closed symmetry algebra with a divergent central charge. In the classical limit, we argue that the algebra may be realized as a geometric symmetry of the hypersurface at future null infinity.  }
		\noindent
   We reduce the massless scalar field theory in Minkowski spacetime to future null infinity. We compute the Poincar\'e flux operators, which can be generalized and identified as the supertranslation and superrotation generators. These generators are shown to form a closed symmetry algebra with a divergent central charge. In the classical limit, we argue that the algebra may be interpreted as the geometric symmetry of a Carrollian manifold, i.e., the hypersurface of future null infinity. Our method may be used to find more physically interesting Carrollian field theories.
  \end{abstract}%We study $(m)$-type connected correlation function(CCF) of OPE blocks with respect to one spatial region in higher dimensional conformal field theory (CFT). The leading term of CCF obeys area law. In the sub-leading terms of CCF, we find logarithmic correction which is cutoff independent.  The logarithmic behaviour is characterized by a parameter $q$ and a constant $p_q$, where $q$ is the maximal power of logarithmic divergence and $p_q$ is the coefficient before the logarithmic term. We derive a UV/IR formula which relates $(m)$-type CCF to $(m-1,1)$-type. A cyclic identity for the coefficient $p_q$ has also been checked carefully.}  	\end{abstract}
	
	%\pacs{04.65.+e,04.70.-s,11.30.-j,12.10.-g}

\end{center}
%%%%%%%%%%%%%%%%%%%%%%%%%%%%%%%%%%%%%%%%%%%%%%%%%%%%%%%%%%%%%%%%%%%%%%%%%%%%%%%%%%%%%%%%

\end{titlepage}
\tableofcontents

\section{Introduction}
The detection of gravitational waves \cite{LIGOScientific:2016aoc}  opens a new window on the observation of the universe. The gravitational wave is one of the greatest predictions of Einstein's equation. Theoretically, it has been known for a long time that the gravitational waves are radiated to future null infinity ($\mathcal{I}^+$) in asymptotically flat spacetime and they transform in the solution space according to the Bondi-Metzner-Sachs (BMS) group \cite{Bondi:1962px, Sachs:1962wk,Sachs:1962zza}. Classically, the BMS group is a semi-direct product of Lorentz group and supertranslations. Over the past decade, there have been various approaches on the understanding of the BMS group. 

The conventional approach is the so-called asymptotic symmetry analysis. By imposing fall off boundary conditions on the solutions of the gravitational field, the BMS group consists of the large diffeomorphisms that preserve the boundary conditions. %In BMS group, The generator of the Lorentz rotations corresponds to global conformal Killing vectors of the celestial sphere while the generator of the supertranslations are smooth functions of the 
The BMS group allows various extensions by including the so-called superrotations. The Barnich-Troessaert (BT) superrotations are  generated by local conformal Killing vectors of the celestial sphere \cite{Barnich:2010eb, Barnich:2009se, Barnich:2010ojg, Barnich:2011mi}. On the other hand, the Campiglia-Laddha (CL) superrotations  are generated by diffeomorphisms of the celestial sphere \cite{ Campiglia:2014yka, Campiglia:2015yka}. Both of them are discussed extensively in the literature.  

The amplitude approach is motivated by the discovery of a set of infrared equivalences \cite{Strominger:2013jfa,Strominger:2017zoo}. Such equivalences relate the BMS asymptotic symmetries, soft theorems \cite{Weinberg:1965nx} and classical memory effects \cite{Zeldovich:1974gvh, Pasterski:2015tva, Nichols:2017rqr, Nichols:2018qac}. As an attempt to apply the holographic principle to flat spacetime, the amplitude approach is to map the S-matrix to conformal correlators living on the celestial sphere \cite{Kapec:2016jld,Pasterski:2016qvg,Pasterski:2017kqt,Raclariu:2021zjz,Pasterski:2021rjz,Donnay:2022aba,Bagchi:2022emh}.

The Carroll group approach is based on the symmetry of the Carroll manifold \cite{Une,Gupta1966OnAA,Henneaux:1979vn}. As is well known,
 the Galilei group could be obtained from the non-relativistic limit (the speed of light $c\to \infty$) of the Poincar\'e group. On the other hand, the Carroll group is the ultra-relativistic limit ($c\to 0$) of the Poincar\'e group, which is the dual of the Galilei group. The BMS group has been shown to be the so-called conformal Carroll group of level 2 \cite{Duval_2014a,Duval_2014b,Duval:2014uoa}. From the point of view of flat holography, it would be interesting to construct field theories with  Carrollian symmetry \cite{Bagchi:2010zz,Bagchi:2016bcd,Bagchi:2019xfx,Bagchi:2019clu,Banerjee:2020qjj,Henneaux:2021yzg,Bagchi:2022owq,Bagchi:2022xug}.
%it is an attempt to apply the holographic principle to flat spacetimes. The key object in this approach is the scattering amplitude.
%Recently, supertranslations and superrotations are the key  for the discovery of a set of infrared equivalences \cite{Strominger:2013jfa,Strominger:2017zoo}.  

In this work, we obtain a scalar field theory by projecting  massless scalar field theory in flat spacetime to its conformal boundary $\mathcal{I}^+$. By imposing the fall-off condition of the scalar field near $\mathcal{I}^+$, we may solve the bulk equation of motion (EOM) asymptotically. There is no constraint on the radiation degree of freedom at the leading order of the EOM. Nevertheless, they form the radiation phase space and obey standard commutation relations in the sense of Ashtekar \cite{1978JMP....19.1542A, Ashtekar:1981bq, Ashtekar:1981sf, Ashtekar:1987tt}. We can define flux operators at $\mathcal{I}^+$ by computing the outgoing Poincar\'e fluxes from radiation. The energy-momentum flux operators are shown to form a Virasoro algebra. By including the angular momentum and the  center-of-mass flux operators, we find a new group which may be regarded as a generalization of the Newman-Unti group of the Carroll manifold $\mathcal{I}^+$. In the soft limit, this new group is reduced to the BMS group.%We could find the BMS group in the soft limit. 
	
This paper is organized as follows. In section \ref{reviewform} we review the BMS group and introduce the conventions used in this work. In section \ref{fluxessec}, we construct the ten Poincar\'e  fluxes radiated to $\mathcal{I}^+$. We compute the commutation relations at $\mathcal{I}^+$ in the following section. In section \ref{sasec}, we compute the commutators of the flux operators and find a closed algebra. We also discuss antipodal matching condition in this section. In section \ref{nusec}, we obtain the same algebra by generalizing the Newman-Unti group of the Carroll manifold $\mathcal{I}^+$. We conclude in section \ref{dissec}. Several technical computations, the derivation of commutators using symplectic structure, and a review about light-ray operator formalism are relegated to four appendices.

\section{Review of the formalism} \label{reviewform}
%In this section, we will first review the BMS group and {\liu define general supertranslations and superrotations.} %Then we will treat the BMS group as a conformal symmetry of the Carroll manifold $\mathcal{I}^+$. 
%\subsection{BMS group}
In Minkowski spacetime $\mathbb{R}^{1,3}$, the metric can be written as 
\bea 
ds^2=-dt^2+dx^idx^i,\quad i=1,2,3.
\eea 
To study radiation at future null infinity $\mathcal{I}^+$,  we can use the retarded coordinate $(u,r,\theta,\phi)$  
\be 
u=t-r,\quad r=\sqrt{\bm x^2},\quad \bm x=(x^1,x^2,x^3),\label{retarded}
\ee and write the metric as 
\bea 
ds^2=-du^2-2du dr+r^2d\Omega^2, 
\eea where 
\bea 
d\Omega^2=d\theta^2+\sin^2\theta d\phi^2\equiv \gamma_{AB}d\theta^Ad\theta^B,\quad A,B=1,2
\eea is the metric of the unit sphere
\bea 
\gamma_{AB}=\left(\begin{array}{cc}1&0\\0&\sin^2\theta\end{array}\right).
\eea In this paper, the covariant derivative $\nabla_A$ is  adapted to the metric $\gamma_{AB}$. 
$\mathcal{I}^+$ can be approached by setting $r\to \infty$ while keeping $u$ fixed.  It has the topology $\mathbb{R}\times S^2$ and can be described by three coordinates 
\bea 
(u,\theta,\phi)=(u,\theta^A).
\eea 

In an asymptotically flat spacetime, the large-$r$ expansion of the metric near $\mathcal{I}^+$ is 
\bea 
ds^2=-du^2-2du dr+r^2\gamma_{AB}d\theta^Ad\theta^B+\delta g_{\mu\nu}dx^\mu dx^\nu.
\eea The original BMS group \cite{Bondi:1962px,Sachs:1962wk} is the large diffeomorphism that preserves the Bondi gauge \be \delta g_{rr}=\delta g_{rA}=0,\quad \partial_r (r^{-4}\det g_{AB})=0 \label{bondigauge}\ee and the boundary fall-off conditions 
\bea 
&&\delta g_{uu}=\mathcal{O}\left(\frac{1}{r}\right),\quad \delta g_{ur}=\mathcal{O}\left(\frac{1}{r^2}\right),\quad \delta g_{uA}=\mathcal{O}(1),\quad \delta g_{AB}=\mathcal{O}(r).\label{falloff}
\eea

Transformations  generated by the vector 
\bea 
\xi_f=f\partial_u+\frac{1}{2}\nabla_A\nabla^A f\partial_r-\frac{\nabla^A f}{r}\partial_A+\cdots\label{GSTs}
\eea are called supertranslations. %Now we will explain the terminology used in this paper.
%\begin{itemize}
%\item Usually, 
The function $f$ is  smooth  on $S^2$. More explicitly, we write it as 
\be 
f=f(\Omega).
\ee %We will call it a special supertranslation (SST).
%\item In this paper, we will also consider the possibility that $f$ is defined on $\mathcal{I}^+$, i.e., it may depend on the retarded time $u$
%\be 
%f=f(u,\Omega).
%\ee  We will call it a general supertranslation(GST).
%\end{itemize} 
Similarly, the transformations  generated by the vector 
\bea 
\xi_Y=\frac{1}{2}u\nabla_AY^A\partial_u-\frac{1}{2}(u+r)\nabla_AY^A\partial_r+(Y^A-\frac{u}{2r}\nabla^A\nabla_BY^B)\partial_A+\cdots\label{GSRs}
\eea  are called superrotations. We will distinguish two cases:
\begin{itemize}
\item The vector $Y^A$ is a  local (singular) conformal Killing vector (CKV) on $S^2$.  In this case, the vector $Y^A$ can be divided into  holomorphic and anti-holomorphic parts. This is the superrotation in the BT sense. We will not consider this case in this paper\footnote{They may relate to cosmic string defects \cite{Compere:2016jwb,Strominger:2016wns}.}. 
\item The vector $Y^A$ is smooth on $S^2$ and generates a diffeomorphism on $S^2$, namely
\be 
Y^A=Y^A(\Omega).
\ee This is the CL superrotation. %We will call it a special superrotation (SSR).
%\item Once $Y^A$ is defined on $\mathcal{I}^+$,  we will call it a general superrotation(GSR).
\end{itemize} 
By combining supertranslations and superrotations, the usual BMS group is generated by the vector 
\be 
\xi_{f,Y}=Y^A(\Omega)\partial_A+[f(\Omega)+\frac{u}{2}\nabla_A Y^A(\Omega)]\partial_u\label{bmsge}
\ee at future null infinity.
%We should emphasize that only SSTs and SSRs are asymptotic symmetries in the previous boundary condition. For GSTs and GSRs generated by \eqref{GSTs} and \eqref{GSRs}, the fall-off conditions \eqref{falloff}  are not preserved. One need to relax the fall-off conditions. Related discussions can be found in  \cite{Adami:2020amw, Adami:2021nnf} for general null hypersurfaces. In this paper, we will not focus on the relaxed fall-off conditions in gravitational theory. Instead, we will show that the vectors \eqref{GSTs} and \eqref{GSRs} may generate large diffeomorphisms in the scalar theory at $\mathcal{I}^+$. 
%\subsection{Conformal Carroll group and Newman-Unti group}

\section{Fluxes}\label{fluxessec}
\begin{figure}
	\centering
	\includegraphics[width=3in]{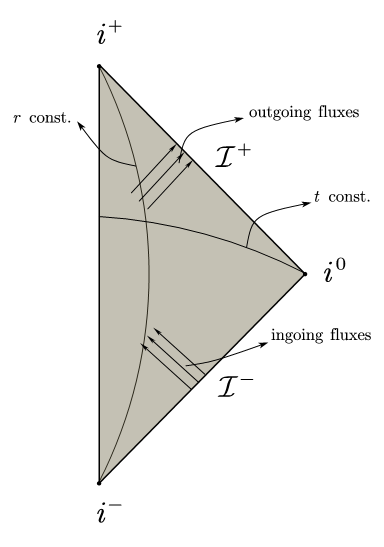}
	\caption{Penrose diagram of the Minkowski spacetime. Massive particles come from past timelike infinity $i^-$ and go to future timelike infinity $i^+$. Massless particles start from past null infinity $\mathcal{I}^-$ and move to future null infinity $\mathcal{I}^+$. $\mathcal{I}^+$ and $\mathcal{I}^-$ are null hypersurfaces with topology $\mathbb{R}\times S^2
		$. There are ingoing fluxes at $\mathcal{I}^-$ and outgoing fluxes at $\mathcal{I}^+$. }
	\label{pendiagram}
\end{figure}
Since BMS symmetry relates to the radiation phase space at $\mathcal{I}^+$, we 
will use a massless real scalar  to study the radiation at $\mathcal{I}^+$.  The action is 
\bea 
S=\int d^4x  [-\frac{1}{2}\partial_\mu\Phi \partial^\mu\Phi-V(\Phi)+J\Phi]. \label{action}
\eea The first term is the kinematic term and the second term is the potential. Since the theory is massless, the potential  is $V(\Phi)\sim \Phi^3$ perturbatively. 
To be more precisely, we may expand the potential as 
\be 
V(\Phi)=\sum_{k=3}^\infty \frac{\lambda_k}{k}\Phi^k. \label{expansionV}
\ee The last term in the action is a source coupled to the field and it causes the scalar radiation. 
The stress-energy tensor of the theory is 
\bea 
T_{\mu\nu}=\partial_\mu\Phi \partial_\nu \Phi+ \eta_{\mu\nu}\mathcal{L},\quad \mu,\nu=0,1,2,3. \label{stress}
\eea where the Lagrangian can be read out from the action \eqref{action}. 

Figure \ref{pendiagram} is the Penrose diagram of the Minkowski spacetime. The scalar theory is in the bulk of the Penrose diagram. At the boundary, there are non-trivial outgoing fluxes at $\mathcal{I}^+$ and ingoing fluxes at $\mathcal{I}^-$. To find the radiation degree of freedom, we should reduce the field $\Phi$ to $\mathcal{I}^+$. This is achieved by imposing the following fall-off condition
\bea 
\Phi&=&\frac{\Sigma(u,\Omega)}{r}+\sum_{k\ge 2}\frac{\Sigma^{(k)}(u,\Omega)}{r^k}\nn\\&=&\frac{\Sigma(u,\Omega)}{r}+\frac{\Sigma^{(2)}(u,\Omega)}{r^2}+\mathcal{O}\left(\frac{1}{r^3}\right)\label{bc}
\eea near $\mathcal{I}^+$.  
Just as in electrodynamics \cite{1998clel.book.....J}, the field $\Sigma$ encodes the radiation degree of freedom. The fields $\Sigma^{(k)},\ k\ge 2$ are subleading terms which will be discussed in the equation of motion. 
The variation of the scalar field under a general diffeomorphism is 
\be 
\delta_\xi\Phi=\xi^\mu\partial_\mu\Phi.
\ee Therefore, the transformation of the boundary field $\Sigma$ under supertranslation is 
\bea 
\delta_f\Sigma&=&f\dot\Sigma \label{sts}
\eea where $\dot\Sigma=\partial_u \Sigma$.
Similarly,  we find the transformation of the field $\Sigma$ under superrotation
\bea 
\delta_Y\Sigma&=&\frac{1}{2}u\nabla_AY^A\dot\Sigma+Y^A\nabla_A\Sigma+\frac{1}{2}\nabla_AY^A \Sigma.\label{srs}
\eea 
The first term on the right hand side of \eqref{srs} may be subtracted,  since it has the same form as the right hand side of \eqref{sts}. For the remaining two terms, we may define %regarded as a supertranslation, we may subtracted it by defining 
\be 
\Delta(Y;\Sigma;u,\Omega)=Y^A\nabla_A\Sigma(u,\Omega)+\frac{1}{2}\nabla_AY^A \Sigma(u,\Omega)\label{DeltaYSig}
\ee for later convenience.

From the action \eqref{action}, the equation of motion is 
\bea  
\Box\Phi-\frac{\partial V}{\partial\Phi}+J=0.
\eea 
The external source affects the vacuum state of the theory and modifies the quantum expectation value and the correlation functions of the field $\Sigma$. 
 However, we will consider quantum fluctuations of the field around the vacuum state, so it is safe to set it to zero.   From now on, we ignore the source term\footnote{In section \ref{antipodalsec}, we will insert back the source term to discuss the antipodal matching condition.} and try to solve  the equation of motion  near $\mathcal{I}^+$. Using the fall-off condition \eqref{bc} and the potential \eqref{expansionV}, we can solve the equation of motion order by order. More explicitly, we have
 \begin{align}
  &\partial_0^2\Phi= \frac{\ddot{\Sigma}}{r}+\frac{\ddot{\Sigma}^{(2)}}{r^2}+\frac{\ddot{\Sigma}^{(3)}}{r^3}+\cdots,
\end{align}
and
\begin{align}
  \partial^2_i\Phi
  =&\frac{\ddot{\Sigma}}{r}+\frac{\ddot\Sigma^{(2)}}{r^2}+\frac{1}{r^3}\bigg[\gamma^{AB}\nabla_A\nabla_B\Sigma+2\dot\Sigma^{(2)}+\ddot{\Sigma}^{(3)}\bigg]+\cdots.
\end{align}
Therefore, we could obtain the following results.
\begin{itemize}
    \item At the leading order $\mathcal{O}(r^{-1})$, the time derivative of $\Sigma$ is cancelled out and there is no corresponding EOM for $\Sigma$. There would  exist term with time derivative of $\Sigma$ from $\partial_0^2\Phi$. However, the contribution about such term from $-\partial_0^2\Phi$ exactly cancels that from $\partial_i^2\Phi$. 
    \item At the subleading order $\mathcal{O}(r^{-2})$, we find the following equation 
    \bea 
    \lambda_3\Sigma^2=0.\label{r-2eom}
    \eea 
     From \eqref{r-2eom}, the field $\Sigma$ should be zero when $\lambda_3\not=0$. To find non-trivial phase space, we may choose $\lambda_3=0$. Then  the equation is valid without imposing any constraint on the field $\Sigma$. 
    \item At next order $\mathcal{O}(r^{-3})$, we find 
    \be 
    -\lambda_4\Sigma^3+2\dot\Sigma^{(2)}+\gamma^{AB}\nabla_A\nabla_B\Sigma=0.\label{sigma2}
    \ee The subleading term $\Sigma^{(2)}$ is determined when we impose the initial condition at the initial time $u=u_i$
    \be 
    \Sigma^{(2)}(u=u_i,\Omega)=\sigma^{(2)}(\Omega).
    \ee  The initial data $\sigma^{(2)}(\Omega)$ is independent of the field $\Sigma$. Equation \eqref{sigma2} indicates that $\Sigma^{(2)}$ is not an independent propagating degree of freedom at $\mathcal{I}^+$.
    \item At higher orders $\mathcal{O}(r^{-k}),\ k\ge 4$, we can also prove that $\Sigma^{(k-1)}$ are  not independent propagating degrees of freedom.  
\end{itemize}
In this paper, the Poincar\'e fluxes could be expressed in terms of the field $\Sigma$ {without any contribution from the higher order terms $\Sigma^{(k)},\ k\ge 2$}. We will focus on field $\Sigma$ in the following. 
% classically. In this paper, we always set $\lambda_3=0$. Then 
%\bea 
%\dot\Sigma^{(2)}&=&-\frac{1}{2}\gamma^{AB}\partial_A\Sigma\partial_B\Sigma+\frac{\lambda_4}{2}\Sigma^3,\\
%\dot\Sigma^{(3)}&=&-\frac{1}{4}\gamma^{AB}\partial_A\Sigma\partial_B\Sigma-\frac{1}{2}\Sigma^{(2)}+\frac{3}{4}\lambda_4\Sigma^2\Sigma^{(2)}.
%\eea The Lagrangian is asymptotically 
%\bea 
%\mathcal{L}=-\frac{\Sigma\dot\Sigma}{r^3}+\frac{1}{r^4}[-2\dot\Sigma\Sigma^{(2)}-\frac{1}{2}\Sigma^2-\Sigma\dot\Sigma^{(2)}-\frac{1}{2}\gamma^{AB}\partial_A\Si

\subsection{Energy and momentum fluxes}
To find the energy flux at $\mathcal{I}^+$, we use the conservation of the stress-energy tensor\footnote{Note that the conservation law is written in Cartesian coordinates, although arguments of terms are in retarded frame. This leads to standard definitions of energy, momentum and angular momentum.}
\be 
\partial_\mu T^{\nu\mu}=0.
\ee It follows that the energy flux can be written as
\bea 
\frac{dP^0}{dt}&=&\frac{d}{dt}\int_V d^3\bm x T^{00}=-\int_V d^3\bm x \partial_i T^{0i}=-\int_{\partial V}dS_i T^{0i}. \label{dp0dt}
\eea 
$T^{0i}$ is the $i$-th component of the energy flux density \cite{2018grav.book.....M}  which is radiated out of surface of the volume $V$. Near $\mathcal{I}^+$, it is expanded as 
\be 
T^{0i}=\frac{\dot\Sigma^2 n^i}{r^2}+\mathcal{O}\left(\frac{1}{r^3}\right). 
\ee The leading term cancels the factor $r^2$ in the integration measure 
\be 
dS_i=r^2 n_i d\Omega
\ee  and leads to a finite result. Note that the right hand side of \eqref{dp0dt} is a function of $u$. In the retarded coordinates \eqref{retarded}, the derivative with respect to $u$ can be found by using the chain rule,  
\be 
\frac{d}{du}=\frac{\partial t}{\partial u}\frac{d}{dt}=\frac{d}{dt}.
\ee 
Therefore,  we find the energy flux 
\bea 
\frac{dP^0}{du}=-\int_{S^2}d\Omega \ \dot{\Sigma}^2(u,\Omega).
\eea In a similar way, we find the following momentum flux 
\bea 
\frac{dP^i}{du}=-\int_{S^2}d\Omega \ \dot{\Sigma}^2(u,\Omega) n^i,
\eea where $n^i$ is the normal vector of the unit sphere $S^2$
\bea 
n^i=(\sin\theta\cos\phi,\sin\theta\cos\phi,\cos\theta).
\eea The energy and momentum radiated to $\mathcal{I}^+$ during the time $(-\infty,u_0)$ can be written as 
\bea 
T_f(u_0)=\int_{-\infty}^{u_0} du d\Omega f(\Omega) \dot{\Sigma}^2, \quad -\infty<u_0<\infty, \label{radiation}
\eea where $f$ is a function on the sphere. 
\begin{itemize}
\item When $f=-1$,  \eqref{radiation} is the energy radiated to $\mathcal{I}^+$. 
\item When $f=-n^i$, \eqref{radiation} is the momentum radiated to $\mathcal{I}^+$. 
\item It is natural to generalize the function $f$ to be any smooth function on the sphere. As we will see later, this corresponds to the supertranslation exactly. 
\end{itemize}
The energy and momentum radiated to $\mathcal{I}^+$ are encoded in the expression \eqref{radiation}. It is easy to see that  \eqref{radiation} can be  naturally generalized to the following version 
\bea 
\mathcal{T}_f=\int_{-\infty}^\infty du d\Omega f(u,\Omega)\dot\Sigma^2.\label{lf}
\eea Now $f$ is a function on $\mathcal{I}^+$. When it is a step function 
\be 
f(u,\Omega)=\theta(u_0-u)g(\Omega),
\ee  $\mathcal{T}_f$ is equivalent to $T_g(u_0)$. If we want to study the time-dependence of the radiation, it is necessary to consider this generalization.  Otherwise, after averaging over time in \eqref{lf}, any time-dependent information of $\dot\Sigma^2(u,\Omega)$ will get lost.

Due to the topology of $\mathcal{I}^+$, 
the function $f(u,\Omega)$ can be expanded in the following basis
\bea 
f(u,\Omega)=\sum_{\ell,m}\int_{-\infty}^\infty d\omega c_{\omega,\ell,m}f_{\omega,\ell,m},\quad f_{\omega,\ell,m}=e^{-i\omega u}Y_{\ell,m}(\Omega),
\eea  where $\omega$ is the frequency which is dual to the retarded time and $(\ell,m)$ are used to label the spherical harmonics. Therefore, \eqref{lf} can also be labeled by three quantum numbers $(\omega,\ell,m)$
\bea 
\mathcal{T}_{\omega,\ell,m}=
\int_{-\infty}^\infty du d\Omega f_{\omega,\ell,m}\dot\Sigma^2.\label{lf2}
\eea 
 As we will show in Appendix \ref{Light-ray}, the insertion of $e^{-i\omega u}$ is also performed in the context of light-ray operator formalism. At the end of this subsection, we will define an energy flux density operator at $\mathcal{I}^+$
\bea 
T(u,\Omega)=\dot\Sigma^2.
\eea It encodes equivalent information of the smeared operator  \eqref{lf}.

\subsection{Angular momentum and  center-of-mass fluxes}
The system is also Lorentz invariant. Thus we can also find a conserved current 
\bea 
M^{\mu\nu\rho}=x^\mu  T^{\nu\rho}-x^\nu T^{\mu\rho},\quad \partial_\rho M^{\mu\nu\rho}=0
\eea 
 whose corresponding conserved charges are the angular momentum and center-of-mass.
Therefore, we can find the following fluxes at $\mathcal{I}^+$
\bea 
\frac{dL^{\mu\nu}}{du}&=&-\int_{V}d^3\bm x\partial_i M^{\mu\nu i}=-\int_{\partial V} dS_i M^{\mu\nu i}.
\eea 
\begin{itemize}
\item For the rotation symmetry, we find the angular momentum fluxes 
\bea 
\frac{dL^{ij}}{du}&=&\int_{S^2}d\Omega Y_{ij}^A \ \dot\Sigma \nabla_A\Sigma, \label{lij}
\eea where $Y_{ij}^A$ are the three Killing vectors of $S^2$ 
\bea 
Y_{12}^A&=&(0,1),\\
Y^A_{23}&=&(-\sin\phi,-\cot\theta\cos\phi),\\
Y^A_{13}&=&(-\cos\phi,\cot\theta\sin\phi).
\eea 
\item For the Lorentz boost, we find the center-of-mass fluxes 
\bea 
\frac{dL^{0i}}{du}&=&-\int_{S^2}d\Omega (Y_i^A\dot\Sigma\nabla_A\Sigma+u \dot\Sigma^2 n^i),\label{l0i}
\eea where $Y_i^A$ are the three strictly conformal Killing vectors of $S^2$
\bea 
Y_1^A&=&(-\cos\theta\cos\phi,\frac{\sin\phi}{\sin\theta}),\\Y_2^A&=&(-\cos\theta\sin\phi,-\frac{\cos\phi}{\sin\theta}),\\ Y_3^A&=&(\sin\theta,0).
\eea They are related to the normal vector $n^i$ by 
\bea 
\nabla_AY^A_i=2n^i.
\eea More properties on the six conformal Killing vectors $Y_i^A, Y_{ij}^A$ can be found in Appendix \ref{ckvs}.
\end{itemize}
From \eqref{lij} and \eqref{l0i}, we can define a smeared  operator
\be
\mathcal{R}_Y=\int_{-\infty}^\infty  du d\Omega\ Y^A(u,\Omega)\dot\Sigma\nabla_A\Sigma,
\ee where $Y^A(u,\Omega)$ is a vector on $\mathcal{I}^+$. 
\begin{itemize}
\item When $Y^A=Y^A_{ij}$, $\mathcal{R}_Y$ is the angular momentum radiated to $\mathcal{I}^+$ during the whole time $(-\infty,\infty)$. 
\item When $Y^A=Y_i^A$, $\mathcal{R}_Y+\mathcal{T}_{\frac{u}{2}\nabla_AY^A}$ is related to  the variation of the center-of-mass during the whole time.
\item The vector $Y^A$ can also be any smooth vector on $S^2$ and  $\mathcal{R}_Y$ will be related to the CL superrotation  \cite{ Campiglia:2014yka}.
\item When $Y^A$ is any singular conformal Killing vector on $S^2$ , $\mathcal{R}_Y$ will be related to the BT superrotation \cite{Barnich:2010eb}.% We will not discuss this choice in this paper, see \cite{Compere:2016jwb, Strominger:2016wns} for more details. 
\item We can choose $Y^A$ to be 
\be 
Y^A(u,\Omega)=\theta(u_0-u)Y^A(\Omega).
\ee Then $\mathcal{R}_Y$ becomes
\bea
R_Y(u_0)=\int_{-\infty}^{u_0}du d\Omega Y^A(\Omega) \dot\Sigma\nabla_A\Sigma.
\eea This is the superrotation charge radiated to $\mathcal{I}^+$ during the time $(-\infty,u_0)$.
\end{itemize}
For later convenience, we  also define an angular momentum flux density operator
\be 
R_A(u,\Omega)=\dot\Sigma\nabla_A\Sigma.
\ee
Actually, we could define a family of such angular momentum flux density operators
\begin{align}
    M_A(\lambda;u,\Omega)=\lambda \dot \Sigma \nabla_A\Sigma-(1-\lambda)\Sigma \nabla_A\dot \Sigma \label{lamMA}
\end{align} where  $\lambda$ is any real constant.
For the cases in which $Y^A$ is independent of $u$, all the operators in this family are equivalent, since we can integrate by parts. In other words, the corresponding smeared operator
\begin{align}
    \mathcal{M}_Y(\lambda)=\int dud\Omega Y^A(\Omega)M_A(\lambda;u,\Omega)
\end{align}
does not depend on $\lambda$.  However, when $Y^A$ is time-dependent, the smeared operators 
\be 
\mathcal{M}_Y(\lambda)=\int du d\Omega Y^A(u,\Omega)M_A(\lambda;u,\Omega)
\ee are not equivalent to each other. 
%There is another density
%\bea 
%\widetilde{R}_A=\Sigma\nabla\dot\Sigma
%\eea 

\section{Canonical quantization}\label{canosec}
In the previous section, we find the Poincar\'e fluxes at $\mathcal{I}^+$.  The densities 
$T(u,\Omega), R_A(u,\Omega)$ are classical objects so far. In this section, we will find the radiative Hilbert space using canonical quantization. The densities $T(u,\Omega),R_A(u,\Omega)$ will become quantum flux density     operators defined on $\mathcal{I}^+$.
\subsection{Commutators}\label{comuscip}
In perturbative quantum field theory, the scalar field $\Phi$ may be quantized  asymptotically using annihilation and creation operators $b_{\bm k}, b^\dagger_{\bm k}$ 
\bea
\Phi(t,\bm x)&=&\int \frac{d^3\bm k}{(2\pi)^3}\frac{1}{\sqrt{2\omega_{\bm{k}}}}(e^{-i\omega t+i\bm{k}\cdot\bm{x}}b_{\bm{k}}+e^{i\omega t-i\bm{k}\cdot\bm{x}}b_{\bm{k}}^\dagger),\label{asymquan}
\eea with the standard commutation relations
\bea 
\ [b_{\bm k},b_{\bm k'}^\dagger]=(2\pi)^3\delta(\bm k-\bm k'),\quad [b_{\bm k},b_{\bm k'}]=0,\quad [b_{\bm k}^\dagger,b_{\bm k'}^\dagger]=0.\label{anncom}
\eea The vector $\bm k$ is the momentum and $\omega$ is the energy. For a massless particle, 
\be 
\omega_{\bm k}=|\bm k|\equiv k.
\ee The plane wave can be expanded as spherical waves  
\bea 
e^{i\bm k\cdot \bm x}=4\pi\sum_{\ell,m}i^\ell j_\ell(\omega r)Y^*_{\ell, m}(\theta,\phi)Y_{\ell, m}(\theta',\phi'),
\eea where the vectors $\bm k$ and $\bm x$ are written in spherical coordinates as 
\bea 
\bm k=(\omega,\theta,\phi),\quad \bm x=(r,\theta',\phi').
\eea The spherical Bessel function $j_\ell(\omega r)$ has the following asymptotic behaviour as $r\to\infty$
\bea 
j_\ell(\omega r)\sim \frac{\sin(\omega r-\frac{\pi\ell}{2})}{\omega r}=\frac{e^{i(\omega r-\pi\ell/2)}-e^{-i(\omega r-\pi\ell/2)}}{2i\omega r}.
\eea Therefore,
\bea 
e^{-i\omega t+i\bm k\cdot\bm x}\sim 4\pi\sum_{\ell m}i^\ell\frac{e^{-i\omega u-i\pi\ell/2}-e^{-i\omega v+i\pi\ell/2}}{2i\omega r}Y^*_{\ell,m}(\theta,\phi)Y_{\ell,m}(\theta',\phi').
\eea Near $\mathcal{I}^+$, the term with $e^{-i\omega v}$ oscillates infinite times and we can set it to zero safely\footnote{It is common to set the term with $e^{-i\omega v}$ to zero in the context of Unruh effect\cite{Crispino:2007eb,Takagi:1986kn}.}. Note that this is also the requirement of the boundary condition \eqref{bc}. We read the quantum version of 
$\Sigma(u,\Omega)$ as 
 \bea 
\Sigma(u,\Omega)&=&\int_0^\infty \frac{d\omega}{\sqrt{4\pi\omega}}\sum_{\ell m}[a_{\omega,\ell,m}e^{-i\omega u}Y_{\ell,m}(\Omega)+a^\dagger_{\omega,\ell,m}e^{i\omega u}Y^*_{\ell,m}(\Omega)]\label{sigmaexp}
\eea
 where
\bea
a_{\omega,\ell,m}&=&\frac{\omega}{2\sqrt{2}\pi^{3/2} i}\int d\Omega b_{\bm k}Y_{\ell,m}^*(\Omega),\label{annihi}\\
a_{\omega,\ell,m}^\dagger&=&\frac{\omega i}{2\sqrt{2}\pi^{3/2}}\int d\Omega b^\dagger_{\bm k}Y_{\ell,m}(\Omega).\label{create}
\eea 
We can also inverse \eqref{annihi} and \eqref{create} as 
\bea
b_{\bm k}&=&\frac{2\sqrt{2}\pi^{3/2}i}{\omega}\sum_{\ell,m}a_{\omega,\ell,m}Y_{\ell,m}(\Omega),\\
b_{\bm k}^\dagger&=& -\frac{2\sqrt{2}\pi^{3/2}i}{\omega}\sum_{\ell,m}a^\dagger_{\omega,\ell,m}Y^*_{\ell,m}(\Omega).
\eea 
We find the following commutators
\bea 
&& [a_{\omega,\ell,m},a_{\omega',\ell',m'}]=[a_{\omega,\ell,m}^\dagger,a_{\omega',\ell',m'}^\dagger]=0,\\
&& [a_{\omega,\ell,m},a_{\omega',\ell',m'}^\dagger]=\delta(\omega-\omega')\delta_{\ell,\ell'}\delta_{m,m'}.
\eea Therefore, $a_{\omega,\ell,m}$ are annihilation operators and $a^\dagger_{\omega,\ell,m}$ are creation operators at $\mathcal{I}^+$.  They are natural operators at $\mathcal{I}^+$ instead of $b_{\bm k}$ and $b_{\bm k}^\dagger$. 

Now we can find the following commutator
\bea 
[\Sigma(u,\Omega),\Sigma(u',\Omega')]&=&\delta(\Omega-\Omega')[\int_{0}^\infty \frac{d\omega}{4\pi\omega} e^{-i\omega(u-u')}-\int_0^\infty \frac{d\omega}{4\pi\omega}e^{i\omega(u-u')}]\nn\\&=&\delta(\Omega-\Omega')\int_{-\infty}^\infty \frac{d\omega}{4\pi\omega}e^{-i\omega(u-u')}.
\eea The Dirac delta function on the sphere is 
\be 
\delta(\Omega-\Omega')=\frac{1}{\sin\theta}\delta(\theta-\theta')\delta(\phi-\phi').\label{diracsphere}
\ee The integral in the commutator is divergent in general. However, we can first compute 
\bea 
\ [\Sigma(u,\Omega),\dot\Sigma(u',\Omega')]&=&\frac{i}{2}\delta(u-u')\delta(\Omega-\Omega'),\label{comsigdot}
\eea and then integrate on $u'$
\bea
[\Sigma(u,\Omega),\Sigma(u',\Omega')]&=&\frac{i}{2}\alpha(u-u')\delta(\Omega-\Omega').\label{comsigma}
\eea The function $\alpha(u-u')$ should satisfy the following two properties
\bea 
\frac{d}{du'}\alpha(u-u')=\delta(u-u'),\quad \alpha(u-u')=-\alpha(u'-u).
\eea These fix the expression of $\alpha(u-u')$
\bea 
\alpha(u-u')=\frac{1}{2}[\theta(u'-u)-\theta(u-u')]
\eea 
where $\theta(x)$ is the Heaviside step function 
\be 
\theta(x)=\left\{\begin{array}{cc}1,&\ x>0\\
0,&\ x<0.\end{array}\right.
\ee 
Finally, we write down the commutator between two $\dot\Sigma$ operators 
\bea 
\ [\dot \Sigma(u,\Omega),\dot\Sigma(u',\Omega')]=\frac{i}{2}\delta'(u-u')\delta(\Omega-\Omega'),\label{comdotdot}
\eea where 
\bea 
\delta'(u-u')\equiv \frac{d}{du}\delta(u-u')=-\frac{d}{du'}\delta(u-u').
\eea 
%\subsection{Asymptotic quantization}
\subsection{Correlation functions}
To compute correlation functions, we need to define the vacuum state. 
Since $a_{\omega,\ell,m}$ is a linear combination of $b_{\bm k}$, the free vacuum is still defined as 
\bea 
a_{\omega,\ell,m}|0\rangle=0\quad\Leftrightarrow \quad b_{\bm k}|0\rangle=0.
\eea  
For interacting theory, the physical vacuum $|\bm 0\rangle$ is not exactly the free vacuum. The physical vacuum is defined as the lowest energy state of the Hamiltonian $H$. We can expand the free vacuum as the superposition of the eigenstates of the Hamiltonian 
\bea 
|0\rangle=|\bm 0\rangle \langle \bm 0|0\rangle+\sum_{n} |\bm n\rangle \langle \bm n|0\rangle.
\eea  We use $n$ to label the eigenstates of the Hamiltonian 
\be 
H|\bm n\rangle=E_n|\bm n\rangle.
\ee 
The energy of the physical vacuum can be shifted to 0 
\be 
H|\bm 0\rangle=0|\bm 0\rangle=0.
\ee Usually, the energy is assumed to be positive for excited states \cite{1995iqft.book.....P}
\be 
E_n>0.
\ee The physical vacuum can be found from
\bea 
\lim_{t\to\infty(1-i\epsilon)}e^{-iHt}|0\rangle=|\bm 0\rangle \langle \bm 0|0\rangle.
\eea In this paper, we will only focus on the theory at $\mathcal{I}^+$. It corresponds to the final states after scattering process. 
The radiative Hilbert space may be constructed by the creation operators acting on the free  vacuum state. For example, a particle with momentum $\bm k=(\omega,\Omega)$ is
a superposition state\bea 
- \frac{4\pi^{3/2}i}{\sqrt\omega}\sum_{\ell,m}Y^*_{\ell,m}(\Omega)a^\dagger_{\omega,\ell,m}|0\rangle.
\eea 
We will derive the symmetry algebra at $\mathcal{I}^+$. It is better to consider the free theory at first. In the following, we will use $|0\rangle$  to denote the vacuum state.  
Using the expansion \eqref{sigmaexp},  the vacuum correlation function with odd number of $\Sigma$ is always zero.  The fundamental two-point correlation functions at $\mathcal{I}^+$ are
\bea
\langle 0|\Sigma(u,\Omega)\Sigma(u',\Omega')|0\rangle&=&\int_0^\infty\frac{d\omega}{4\pi\omega}e^{-i\omega(u-u')}\delta(\Omega-\Omega'),\label{sigmasigma}\\
\langle 0|\Sigma(u,\Omega)\dot\Sigma(u',\Omega')|0\rangle&=&i\int_0^\infty\frac{d\omega}{4\pi}e^{-i\omega(u-u')}\delta(\Omega-\Omega'),\label{sigmadot}\\
\langle 0|\dot\Sigma(u,\Omega)\Sigma(u',\Omega')|0\rangle&=&-i\int_0^\infty\frac{d\omega}{4\pi}e^{-i\omega(u-u')}\delta(\Omega-\Omega'),\label{dotsigma}\\
\langle 0|\dot\Sigma(u,\Omega)\dot\Sigma(u',\Omega')|0\rangle&=&\int_0^\infty\frac{d\omega}{4\pi}\omega e^{-i\omega(u-u')}\delta(\Omega-\Omega').
\label{dotdot}\eea 
The integral in \eqref{sigmadot} is only well defined when $\text{Im}(u-u')<0$. Therefore, we use the following $i\epsilon$ prescription \cite{haag:1993,Hartman:2015lfa} in the correlators
\bea 
u\to u-i\epsilon,\quad \epsilon>0.
\eea Now the correlators \eqref{sigmadot}-\eqref{dotdot} are 
\bea 
\langle 0|\Sigma(u,\Omega)\dot\Sigma(u',\Omega')|0\rangle&=&\frac{1}{4\pi(u-u'-i\epsilon)}\delta(\Omega-\Omega'),\label{corsigdot}\\
\langle 0|\dot\Sigma(u,\Omega)\Sigma(u',\Omega')|0\rangle&=&-\frac{1}{4\pi(u-u'-i\epsilon)}\delta(\Omega-\Omega'),\label{cordotsig}\\
\langle 0|\dot\Sigma(u,\Omega)\dot\Sigma(u',\Omega')|0\rangle&=&-\frac{1}{4\pi(u-u'-i\epsilon)^2}\delta(\Omega-\Omega').\label{cordotsigdotsig}
\eea The correlator \eqref{sigmasigma} is still divergent. Nevertheless, we write it as 
\bea 
\langle 0|\Sigma(u,\Omega)\Sigma(u',\Omega')|0\rangle=\beta(u-u')\delta(\Omega-\Omega'), \label{prop}
\eea where
\bea 
\beta(u-u')=\int_0^\infty \frac{d\omega}{4\pi\omega}e^{-i\omega(u-u'-i\epsilon)}.
\eea This $\beta$ function may be regularized by 
\bea 
\beta(u-u')=\lim_{ \kappa\to 0^+}\int_0^\infty \frac{d\omega}{4\pi\omega^{1-\kappa}}e^{-i\omega(u-u'-i\epsilon)}.
\eea Besides a divergent part which is proportional to $\kappa^{-1}$, $\beta(u-u')$ should be a logarithmic function 
\be 
\beta(u-u')\sim \frac{1}{4\pi\kappa}-\frac{1}{4\pi}\log(i(u-u'-i\epsilon))-\frac{\gamma_E}{4\pi},
\ee where $\gamma_E$ is the Euler constant.  There is a branch point at $u=u'$. Though $\beta(u-u')$ is divergent, we find a finite result by considering the following difference
\bea 
\beta(u-u')-\beta(u'-u)=\frac{i}{2}\alpha(u-u').\label{alphabeta}
\eea  This is consistent with the commutator \eqref{comsigma}. The time derivative of the $\beta(u-u')$ function is 
\bea
\frac{d}{du}\beta(u-u')=-\frac{d}{du'}\beta(u-u')=-\frac{1}{4\pi(u-u'-i\epsilon)}.\label{deriv-beta}
\eea We also notice that the correlators \eqref{corsigdot}-\eqref{cordotsigdotsig} are consistent with the commutators \eqref{comsigdot} and \eqref{comdotdot} by using the following formula
\bea 
\frac{1}{x\pm i\epsilon}=P\frac{1}{x}\mp i\pi\delta(x)
\eea where $P\frac{1}{x}$ is the principal value.
Now we compute the four-point correlation function using Wick contraction 
\bea 
\langle 0|\Sigma(u_1,\Omega_1)\Sigma(u_2,\Omega_2)\Sigma(u_3,\Omega_3)\Sigma(u_4,\Omega_4)|0\rangle&=&G_{12}G_{34}+G_{13}G_{24}+G_{14}G_{23},\label{4pt}
\eea where we have defined the two-point functions 
\bea 
G_{ij}=\langle 0|\Sigma(u_i,\Omega_i)\Sigma(u_j,\Omega_j)|0\rangle=\beta(u_i-u_j)\delta(\Omega_i-\Omega_j).
\eea All other four-point correlation functions are generated from \eqref{4pt}. For example, 
\bea 
\langle 0|\dot\Sigma(u_1,\Omega_1)\dot\Sigma(u_2,\Omega_2)\dot\Sigma(u_3,\Omega_3)\dot\Sigma(u_4,\Omega_4)|0\rangle&=&H_{12}H_{34}+H_{13}H_{24}+H_{14}H_{23}, 
\eea where 
\bea 
H_{ij}=\partial_{u_i}\partial_{u_j}G_{ij}=-\frac{1}{4\pi(u_i-u_j-i\epsilon)^2}\delta(\Omega_i-\Omega_j).
\eea 

Attentive readers may have questions about the correlation functions  in this subsection and the commutation relations in the previous subsection.  Usually, the propagator 
\be 
\langle  \Phi(t,\bm x)\Phi(t',\bm x')\rangle \label{greenphi} \ee is understood as the corresponding Green function which satisfies the wave equation. Interestingly, we have concluded that the propagating field $\Sigma$ is not subject to any additional constraint upon $\lambda_3=0$. At the same time, there is still a non-trivial propagator  
\be 
\langle\Sigma(u,\Omega)\Sigma(u',\Omega') \rangle.\label{greensigma}
\ee  
There is no contradiction since the EOM alone does not fix the propagator. One should also impose suitable initial/boundary conditions.  For example, any plane wave 
\be 
e^{-i\omega t+i\bm k\cdot\bm x}
\ee could satisfy the bulk EOM once the energy $\omega$ and the momentum $\bm k$ are related by the identity $\omega=|\bm k|$. Due to the completeness of the Fourier modes in the solution space, the bulk field can be expanded in the plane wave basis with suitable coefficients 
\bea 
\Phi(t,\bm x)&=&\int \frac{d^3\bm k}{(2\pi)^3}\frac{1}{\sqrt{2\omega_{\bm{k}}}}(e^{-i\omega t+i\bm{k}\cdot\bm{x}}b_{\bm{k}}+e^{i\omega t-i\bm{k}\cdot\bm{x}}b_{\bm{k}}^\dagger).
\eea  In canonical quantization, one should impose non-trivial commutators between $\Phi$ and its conjugate momentum $\partial_t\Phi$ which turns into the commutators between $b_{\bm k}$ and its Hermitian conjugate. With the definition of vacuum, one can determine the propagator using the commutators.

There is a similar story for the boundary propagator \eqref{greensigma}. Any smooth function at the boundary may be expanded in the basis $f_{\omega,\ell,m}=e^{-i\omega u}Y_{\ell,m}(\Omega)$ since  boundary manifold is $\mathcal{I}^+=\mathbb{R}\times S^2$.  There is no further constraint for the three quantum numbers $(\omega,\ell,m)$ since there is no dynamical EOM for the boundary field $\Sigma$. This is exactly the mode expansion  \eqref{sigmaexp} of the boundary field which is written below
\bea 
\Sigma(u,\Omega)&=&\int_0^\infty \frac{d\omega}{\sqrt{4\pi\omega}}\sum_{\ell m}[a_{\omega,\ell,m}e^{-i\omega u}Y_{\ell,m}(\Omega)+a^\dagger_{\omega,\ell,m}e^{i\omega u}Y^*_{\ell,m}(\Omega)].
\eea There are still non-trivial commutators between the coefficients $a_{\omega,\ell,m}$ and $a^\dagger_{\omega,\ell,m}$ which are inherited from the quantum bulk field. 
  In other words, though the boundary theory $\Sigma$ has no dynamical EOM, there is indeed a non-trivial symplectic structure in the phase space, from which we could work out the Poisson brackets and hence commutation relations. The derivations are collected in Appendix \ref{hamiltonmethod}. The propagator of the boundary field $\Sigma$ is a consequence of the symplectic structure and the definition of vacuum state.%The commutation relations and propagators are directly calculated from asymptotic mode expansion of scalar field. The canonical quantization procedure is done in the bulk, and we projects the quantum objects to $\mathcal{I}^+$. 

%Therefore $\dot{\Sigma}(u,\Omega)$ can be written as
%\bea
%\dot{\Sigma}(u,\Omega)&=&-i\int_0^\infty d\omega \sqrt{\frac{\omega}{4\pi}}\sum_{\ell m}[a_{\omega,\ell,m}e^{-i\omega u}Y_{\ell,m}(\Omega)-a^\dagger_{\omega,\ell,m}e^{i\omega u}Y^*_{\ell,m}(\Omega)].
%\eea Then we find 
\subsection{Normal ordering}
After quantization, the densities $T(u,\Omega),R_{A}(u,\Omega)$ are operators. We should refine their definition by using normal ordering 
\bea 
T(u,\Omega)&=&:\dot\Sigma^2(u,\Omega):,\\
R_A(u,\Omega)&=&:\dot\Sigma\nabla_A\Sigma(u,\Omega):.
\eea The procedure is to move the annihilation operators to the right of the creation operators. Therefore, the vacuum expectation values of these flux density operators vanish
\bea 
\langle 0|T(u,\Omega)|0\rangle=\langle0|R_A(u,\Omega)|0\rangle=0.
\eea An equivalent way is to refine the operators by taking the following limit 
\bea
T(u,\Omega)&=&\lim_{u'\to u,\Omega'\to\Omega} \dot\Sigma(u,\Omega)\dot\Sigma(u',\Omega')-\langle 0|\dot\Sigma(u,\Omega)\dot\Sigma(u',\Omega')|0\rangle,\\
R_A(u,\Omega)&=&\lim_{u'\to u,\Omega'\to\Omega} \dot\Sigma(u,\Omega)\nabla_{A'}\Sigma(u',\Omega')-\langle 0|\dot\Sigma(u,\Omega)\nabla_{A'}\Sigma(u',\Omega')|0\rangle.
\eea Considering the normal ordering, we find the following two-point functions
\bea 
\langle 0|T(u,\Omega)T(u',\Omega')|0\rangle&=&\frac{\delta^{(2)}(0)}{8\pi^2(u-u'-i\epsilon)^4}\delta(\Omega-\Omega'),\label{TT2pt}\\
\langle 0|T(u,\Omega)R_{A'}(u',\Omega')|0\rangle&=&\frac{\delta^{(2)}(0)}{16\pi^2(u-u'-i\epsilon)^3}\nabla_{A'}\delta(\Omega-\Omega'),\label{TR2pt}\\
\langle 0|R_{A}(u,\Omega)R_{B'}(u',\Omega')|0\rangle&=&-\frac{\beta(u-u')}{4\pi(u-u'-i\epsilon)^2}\delta(\Omega-\Omega')\nabla_A\nabla_{B'}\delta(\Omega-\Omega')\nn\\&&-\frac{1}{16\pi^2(u-u'-i\epsilon)^2}\nabla_A\delta(\Omega-\Omega')\nabla_{B'}\delta(\Omega-\Omega').\label{MM2pt}
\eea The divergent constant $\delta^{(2)}(0)$ is the Dirac functions \eqref{diracsphere}  on the sphere with the argument equalling to 0. The non-vanishing of the two-point function \eqref{TR2pt} indicates that $R_A$ is not orthogonal to $T$. We define a new operator 
\bea 
M_A=\frac{1}{2}(:\dot\Sigma\nabla_A\Sigma-\Sigma\nabla_A\dot\Sigma:).
\eea It is related to $R_A$ by 
\bea 
M_A=R_A-\frac{1}{4}\nabla_A\dot Q,\label{refine}
\eea where the operator $Q$ is  defined as
\bea
Q\equiv:\Sigma^2:.
\eea Then the two-point function between $T$ and $M$ becomes  zero 
\bea
\langle 0|T(u,\Omega)M_{A'}(u',\Omega')|0\rangle=0.\label{tma}
\eea
 The flux density operators defined in \eqref{lamMA} are classically equivalent in the cases where $Y^A$ is independent of time $u$. However, in the quantum cases, only the operator with $\lambda=1/2$ is orthogonal to $T$. Therefore, it is natural to choose such an operator to define our smeared superrotation flux operator. As will be shown later, we obtain a set of concise commutation relations with such a choice. However, one may also choose other values of $\lambda$ since the orthogonality condition is not necessary.

The two-point function between two $M$s is
\bea
\langle 0|M_A(u,\Omega)M_{B'}(u',\Omega')|0\rangle=-\frac{\beta(u-u')-\frac{1}{4\pi}}{8\pi(u-u'-i\epsilon)^2}\times \Lambda_{AB'}(\Omega,\Omega'),\eea  with 
\be 
\Lambda_{AB'}(\Omega,\Omega')=\delta(\Omega-\Omega')\nabla_A\nabla_{B'}\delta(\Omega-\Omega')-\nabla_A\delta(\Omega-\Omega')\nabla_{B'}\delta(\Omega-\Omega').
\ee 
%For later convenience, we define a smeared operator
To simplify notation, we will use the convention 
\bea 
\mathcal{M}_Y\equiv\mathcal{M}_Y(\lambda=\frac{1}{2})=\int du d\Omega\  Y^A(u,\Omega) M_A(u,\Omega).%,\quad \mathcal{Q}_g=\int du d\Omega\  g(u,\Omega) Q(u,\Omega).
\eea

\subsection{Correlation functions between flux operators}
From the correlation functions between flux density operators, it is easy to calculate correlation functions between flux operators. We define the following quantities
\bea 
\mathbb{T}_f(u)=\int d\Omega f(u,\Omega)T(u,\Omega),\quad \mathbb{M}_Y(u)=\int d\Omega Y^A(u,\Omega)M_A(u,\Omega).
\eea 
They are related to the flux operators 
\bea 
\mathcal{T}_f=\int du \mathbb{T}_f(u),\quad \mathcal{M}_Y=\int du \mathbb{M}_Y(u).
\eea 
From the correlation functions of the flux density operators, we could find
\bea 
\langle \mathbb{T}_{f_1}(u)\mathbb{T}_{f_2}(u')\rangle&=&\frac{\delta^{(2)}(0)}{8\pi^2}\int d\Omega \frac{f_1(u,\Omega)f_2(u',\Omega)}{(u-u'-i\epsilon)^4},\label{tfutfu}\\
\langle \mathbb{T}_{f}(u)\mathbb{M}_Y(u')\rangle&=&0,\\ 
\langle \mathbb{M}_Y(u)\mathbb{M}_Z(u')\rangle &=&-\int d\Omega d\Omega' Y^A(u,\Omega)Z^{B'}(u',\Omega')\Lambda_{AB'}(\Omega,\Omega')\frac{\beta(u-u')-\frac{1}{4\pi}}{8\pi(u-u'-i\epsilon)^2}.\label{myumzu}
\eea Thus the correlation functions of the energy flux operators are
\begin{align}   
  \braket{0|\mathcal{T}_{f_1}\mathcal{T}_{f_2}|0}%=&\int dud\Omega du'd\Omega'f_1(u,\Omega)f_2(u',\Omega')\braket{T(u,\Omega)T(u',\Omega')}\nn\\
  %=&\int dud\Omega du'd\Omega'f_1(u,\Omega)f_2(u',\Omega')\frac{\delta^{(2)}(0)\delta(\Omega-\Omega')}{8\pi^2(u-u'-i\epsilon)^4}\nn\\
  =&\frac{\delta^{(2)}(0)}{8\pi^2}\int dud\Omega du'\frac{f_1(u,\Omega)f_2(u',\Omega)}{(u-u'-i\epsilon)^4}.
\end{align}
When $f_1,f_2$ do not depend on $u$, it is easy to see that
\begin{align}
  \braket{0|\mathcal{T}_{f_1}\mathcal{T}_{f_2}|0}
  =&\frac{\delta^{(2)}(0)}{8\pi^2}\int dud\Omega du'\frac{f_1(\Omega)f_2(\Omega)}{(u-u'-i\epsilon)^4}=0.
\end{align}
As a consequence of \eqref{tma}, we find
\begin{align}
    \bra{0}\mathcal{T}_f\mathcal{M}_Y\ket{0}=0.
\end{align}
At last, we consider $\bra{0}\mathcal{M}_Y\mathcal{M}_Z\ket{0}$
\begin{align}   
  \braket{0|\mathcal{M}_{Y}\mathcal{M}_{Z}|0}%=&\int dud\Omega du'd\Omega'Y^A(u,\Omega)Z^{B'}(u',\Omega')\braket{M_A(u,\Omega)M_{B'}(u',\Omega')}\nn\\
  =&-\int dud\Omega du'd\Omega'Y^A(u,\Omega)Z^{B'}(u',\Omega')\frac{\beta(u-u')-\frac{1}{4\pi}}{8\pi(u-u'-i\epsilon)^2}\times \Lambda_{AB'}(\Omega,\Omega').
\end{align}
If $Y^A,Z^A$ do not depend on $u$, we find
\begin{align}   
  \braket{0|\mathcal{M}_{Y}\mathcal{M}_{Z}|0}
  =&-\int dud\Omega du'd\Omega'Y^A(\Omega)Z^{B'}(\Omega')\frac{\beta(u-u')-\frac{1}{4\pi}}{8\pi(u-u'-i\epsilon)^2}\times \Lambda_{AB'}(\Omega,\Omega')\nn\\
  =&\int dud\Omega du'd\Omega'Y^A(\Omega)\Lambda_{AB'}(\Omega,\Omega')Z^{B'}(\Omega')\times\frac{1}{32\pi^2(u-u'-i\epsilon)^2} =0.
\end{align}
To obtain the second line, we have integrated by parts with respect to $u$, and have used the derivative of $\beta(u-u')$ given in \eqref{deriv-beta}. 

In summary, when $f_1,f_2$ and $Y^A,Z^A$ are time-independent, the correlators between the smeared operators are zeros
\be 
\langle\mathcal{T}_{f_1}\mathcal{T}_{f_2}\rangle=\langle\mathcal{T}_f\mathcal{M}_Y\rangle=\langle\mathcal{M}_Y\mathcal{M}_Z\rangle=0.
\ee %for time-independent $f$ and $Y^A$. 
However, we have obtained the correlation functions \eqref{tfutfu} and \eqref{myumzu}. They do not vanish and contain more detailed information than the smeared operators when $f_1,f_2$ and $Y^A,Z^A$ are time-independent
\bea 
\langle \mathbb{T}_{f_1}(u)\mathbb{T}_{f_2}(u')\rangle&=&\frac{\delta^{(2)}(0)}{8\pi^2}\int d\Omega \frac{f_1(\Omega)f_2(\Omega)}{(u-u'-i\epsilon)^4},\\
\langle \mathbb{M}_Y(u)\mathbb{M}_Z(u')\rangle &=&-\int d\Omega d\Omega' Y^A(\Omega)Z^{B'}(\Omega')\Lambda_{AB'}(\Omega,\Omega')\frac{\beta(u-u')-\frac{1}{4\pi}}{8\pi(u-u'-i\epsilon)^2}.
\eea  When $u$ and $u'$ are close to each other, there is a peak for the correlation function $\langle \mathbb{T}_{f_1}(u)\mathbb{T}_{f_2}(u')\rangle$. One can also find a peak for $\langle \mathbb{M}_Y(u)\mathbb{M}_Z(u')\rangle$. In the limit 
\be 
u\to\infty,\quad u'\ \text{finite},
\ee the correlation function $\langle \mathbb{T}_{f_1}(u)\mathbb{T}_{f_2}(u')\rangle$ decays by power
\be 
\langle \mathbb{T}_{f_1}(u)\mathbb{T}_{f_2}(u')\rangle\sim \frac{1}{u^4}
\ee and $\langle \mathbb{M}_Y(u)\mathbb{M}_Z(u')\rangle$ decays as 
\bea 
\langle \mathbb{M}_Y(u)\mathbb{M}_Z(u')\rangle\sim \frac{\log u}{u^2}.
\eea

\section{Symmetry algebra at $\mathcal{I}^+$}\label{sasec}
In this section, we will relate the operators $\mathcal{T}_f$ and $\mathcal{M}_Y$ to supertranslation and superrotation respectively. Then we will generalize the BMS algebra by computing the commutators between these flux operators.
\subsection{Supertranslation and superrotation generators}
%Using the Wick contraction, we find the following OPEs
%\bea 
%T(u,\Omega)\Sigma(u',\Omega')&\sim&-\frac{\dot\Sigma(u',\Omega')}{2\pi(u-u')}\delta(\Omega-\Omega')+\cdots,\\
%T(u,\Omega)\dot\Sigma(u',\Omega')&\sim&-\frac{\dot\Sigma(u',\Omega')}{2\pi(u-u')^2}\delta(\Omega-\Omega')-\frac{\ddot\Sigma(u',\Omega')}{2\pi(u-u')}\delta(\Omega-\Omega')+\cdots.
%\eea The second OPE indicates that $\dot\Sigma$ is a primary field with conformal dimension 1.  Now we can transform the OPEs to the commutator
Using the {commutator} \eqref{comsigdot}, we can find 
\bea 
\bar{\delta}_f\Sigma(u',\Omega')\equiv [\mathcal{T}_f,\Sigma(u',\Omega')]&=&-if(u',\Omega')\dot\Sigma(u',\Omega').
\eea This is exactly the transformation of the field under supertranslation \eqref{sts} up to a constant factor. Therefore, when $f$ is time-independent, the operator $\mathcal{T}_f$ should be regarded as the generator of supertranslation. Interestingly, the test function $f$ in $\mathcal{T}_f$ could be time-dependent.  Now we will explain the terminology used in this paper.
\begin{itemize}
\item Usually, $f$ is a smooth function on $S^2$. More explicitly, we write it as 
\be 
f=f(\Omega).
\ee We will call it a special supertranslation (SST).
\item In this paper, we will also consider the possibility that $f$ is defined on $\mathcal{I}^+$, i.e., it may depend on the retarded time $u$
\be 
f=f(u,\Omega).
\ee  We  call it a general supertranslation (GST). Note that we define the GSTs through the flux operators in the scalar field theory at $\mathcal{I}^+$. They are not the ``real'' supertranslations since a time-dependent function $f$ in \eqref{GSTs} would violate the fall-off conditions \eqref{falloff}. 
\end{itemize} 

Similarly, %we can find the OPE between $M_A$ and $\Sigma$
%\bea 
%M_A(u,\Omega)\Sigma(u',\Omega')&\sim&\frac{1}{4\pi(u-u')}\ [\Sigma(u',\Omega')\nabla_A\delta(\Omega-\Omega')-2\nabla_{A'}\Sigma(u',\Omega')\delta(\Omega-\Omega')]\nn\\&&+\frac{1}{2}\frac{d}{du}\{\beta(u-u')[\Sigma(u,\Omega)\nabla_A\delta(\Omega-\Omega')-\nabla_A\Sigma(u,\Omega)\delta(\Omega-\Omega')]\},
%\eea at the second line, it is a totally derivative.  Then we can read out the following commutator
we find the following commutator
\bea 
\bar{\delta}_Y\Sigma(u',\Omega')&\equiv& [\mathcal{M}_Y,\Sigma(u',\Omega')]\nn\\&=&-iY^{A'}(u',\Omega')\nabla_{A'}\Sigma(u',\Omega')-\frac{i}{2}\nabla_{A'}Y^{A'}(u',\Omega') \Sigma(u',\Omega')\nn\\&&+\frac{i}{2}\int du \alpha(u-u')[\dot{Y}^{A'}(u,\Omega')\nabla_{A'}\Sigma(u,\Omega')+\frac{1}{2}\nabla_{A'}\dot{Y}^{A'}(u,\Omega')\Sigma(u,\Omega')]\nn\\&=&-i\Delta(Y;\Sigma;u',\Omega')+\frac{i}{2}\int du \alpha(u-u') \Delta(\dot{Y};\Sigma;u,\Omega').\label{GSRtransint}
\eea At the last step, we have used the definition of $\Delta(Y;\Sigma;u,\Omega)$ in \eqref{DeltaYSig}.  The transformation of $\Sigma(u',\Omega')$ under  the action of $\mathcal{M}_Y$ contains two parts. The first part is local since it only depends locally on the field with the same time. The commutator \eqref{GSRtransint} can be interpreted as a superrotation transformation when $\dot Y=0$, since compared to the above ${\delta}_Y\Sigma$ \eqref{srs}, the additional non-local term vanishes. Just like supertranslation, we distinguish the following two cases.
\begin{itemize} 
\item The CL superrotation is time-independent and we will call it a special superrotation 
 (SSR).
\item Once $Y^A$ is defined on $\mathcal{I}^+$, namely, 
\be 
Y^A=Y^A(u,\Omega),
\ee we will call it a general superrotation (GSR). GSRs are not ``real'' superrotations since they  not only violate the fall-off condition \eqref{falloff}, but also break the null structure of $\mathcal{I}^+$. The latter point will be discussed in section \ref{sec:Newman-Unti}.
\end{itemize}  % with
%\be 
%{\jl{\Delta(Y;\Sigma;u',\Omega')=Y^{A'}(u',\Omega')\nabla_{A'}\Sigma(u',\Omega')+\frac{1}{2}\nabla_{A'}Y^{A'}(u',\Omega') \Sigma(u',\Omega').}}
 The second part is non-local  which is a superposition of the superrotation transformations $\Delta(\dot Y;\Sigma;u,\Omega')$ at different times. 
When the vector $Y^A$ is time-independent, the non-local term vanishes. In this case, the operator 
\be 
\mathcal{T}_{\frac{1}{2}u\nabla_AY^A}+\mathcal{M}_Y
\ee generates the SSRs \eqref{srs}.  In summary, we find the supertranslation and superrotation generators.
\begin{itemize}
\item Supertranslation generators $\mathcal{T}_f$. It is the smeared operator of the energy flux density operator $T(u,\Omega)$. 
\item Superrotation generators $\mathcal{T}_{\frac{1}{2}u\nabla_AY^A}+\mathcal{M}_Y$. Since the first part $\mathcal{T}_{\frac{1}{2}u\nabla_AY^A}$ is just a supertranslation generator, we may also call $\mathcal{M}_Y$ as a superrotation generator. It is a smeared operator of the angular momentum flux density operator $M_A(u,\Omega)$. 
\end{itemize}

\subsection{Symmetry algebra from flux operators}
Since $\mathcal{T}_f$ and $\mathcal{M}_Y$ are identified as supertranslation generator and superrotation generator respectively, they should form a BMS algebra. It is straightforward to find the following commutators
 \bea 
\ \hspace{-14pt}[\mathcal{T}_{f_1},\mathcal{T}_{f_2}]\hspace{-8pt}&=&\hspace{-8pt}C_T(f_1,f_2)+i\mathcal{T}_{f_1\dot f_2-f_2\dot f_1},\label{virasorosub}\\
\ \hspace{-14pt}[\mathcal{T}_f,\mathcal{M}_Y]\hspace{-8pt}&=&\hspace{-8pt}i\mathcal{M}_{f\dot Y}-i\mathcal{T}_{Y^A\nabla_Af}+\frac{i}{4}\mathcal{Q}_{\frac{d}{du}(\dot{Y}^A\nabla_A f)},\label{nonlm}\\
\ \hspace{-14pt}[\mathcal{M}_Y,\mathcal{M}_Z]\hspace{-8pt}&=&\hspace{-8pt}C_M(Y,Z)+i\mathcal{M}_{[Y,Z]}+\frac{i}{2}\int du du'd\Omega \alpha(u'-u):\Delta(\dot{Y};\Sigma;u',\Omega)\Delta(\dot{Z};\Sigma;u,\Omega):.\label{nonmm}
\eea 
Unfortunately, this is not a standard algebra in general. Besides the local operators on the right hand side of the commutators, many interesting new features appear.
\begin{itemize}
	\item There is a new operator 
	\be 
	\mathcal{Q}_g=\int du d\Omega g(u,\Omega) :\Sigma^2:
	\ee on the right hand side of the commutator between supertranslation and superrotation generators \eqref{nonlm}. This operator is absent for $\dot Y=0$.  The commutator between $\mathcal{Q}_g$ and the field $\Sigma$ is 
	\bea 
	\ [\mathcal{Q}_g,\Sigma(u',\Omega')]=i\int du \alpha(u-u')g(u,\Omega')\Sigma(u,\Omega').
	\eea There is no obvious geometric meaning for this operator. However, we may relate it to the light-ray operator of $\Phi^2$ at $\mathcal{I}^+$. This point is clarified in Appendix \ref{Light-ray}. We should also  include it once we consider GSRs. With this operator, we obtain the following three commutators
 \bea 
 \ \hspace{-24pt}[\mathcal{T}_f,\mathcal{Q}_g]\hspace{-8pt}&=&\hspace{-8pt}C_{TQ}(f,g)+i\mathcal{Q}_{\frac{d}{du}(fg)},\label{nonlq}\\
\ \hspace{-24pt}[\mathcal{M}_Y,\mathcal{Q}_g]\hspace{-8pt}&=&\hspace{-8pt}i\mathcal{Q}_{Y^A\nabla_A g}{+i}\int du du'd\Omega \alpha(u'-u)g(u,\Omega):\Sigma(u,\Omega)\Delta(\dot{Y};\Sigma;u',\Omega):,\label{nonmq}\\
\ \hspace{-24pt}[\mathcal{Q}_{g_1},\mathcal{Q}_{g_2}]\hspace{-8pt}&=&\hspace{-8pt}C_Q(g_1,g_2)+2i\int du du'd\Omega \alpha(u'-u)g_1(u',\Omega)g_2(u,\Omega):\Sigma(u,\Omega)\Sigma(u',\Omega):.\label{nonqq}
 \eea 
	
\item Central charges. There are {two} central extension terms in the algebra \eqref{virasorosub}-\eqref{nonmm}
\bea 
C_T(f_1,f_2)&=&-\frac{i}{48\pi}\delta^{(2)}(0)\mathcal{I}_{f_1\dddot {f_2}-f_2\dddot {f_1}},\\
C_M(Y,Z)&=&\int du du'd\Omega d\Omega' Y^A(u,\Omega)Z^{B'}(u',\Omega')\Lambda_{AB'}(\Omega,\Omega')\eta(u-u').
\eea where 
\be 
\eta(u-u')= -\frac{\beta(u-u')-\frac{1}{4\pi}}{8\pi(u-u'-i\epsilon)^2}+\frac{\beta(u'-u)-\frac{1}{4\pi}}{8\pi(u'-u-i\epsilon)^2}.
\ee 
The identity operator is defined as 
\bea 
\mathcal{I}_f=\int du d\Omega f(u,\Omega).
\eea 
There are also two central extension terms in \eqref{nonlq}-\eqref{nonqq}
\bea 
C_{TQ}(f,g)&=&-\frac{i}{4\pi}\delta^{(2)}(0)\mathcal{I}_{f\dot g},\\
C_Q(g_1,g_2)&=&2\delta^{(2)}(0)\int du du'd\Omega [\beta(u-u')^2-\beta(u'-u)^2] g_1(u,\Omega)g_2(u',\Omega).\label{centralqq}
\eea 
These central terms can be read out from two-point correlation functions of the corresponding operators.
Three of the central extension terms $C_T,C_Q,C_{TQ}$ are divergent due to the Dirac delta function $\delta^{(2)}(0)$ on the sphere. We will use a constant $c$ to denote $\delta^{(2)}(0)$ 
\be 
c=\delta^{(2)}(0)
\ee and  will discuss the regularization of $c$ later.
\item Virasoro algebra in higher dimension.
We notice that \eqref{virasorosub} actually forms a Virasoro algebra. To see this point, we transform the supertranslation generator to its Fourier space using \eqref{lf2}
\bea 
[\mathcal{T}_{\omega,\ell,m},\mathcal{T}_{\omega',\ell',m'}]=(\omega'-\omega)\sum_{L=|\ell-\ell'|}^{\ell+\ell'}\sum_{M=-L}^L c_{\ell,m;\ell',m';L,M}\mathcal{T}_{\omega+\omega',L,M}-\frac{\omega^3}{12}c\ \delta({\omega+\omega'})(-1)^m\delta_{\ell,\ell'}\delta_{m, -m'} .\nn\\ \label{virasoro}
\eea 
The constants $c_{\ell,m;\ell',m';L,M}$ are from the decomposition of the product of two spherical harmonic functions into the summation of spherical harmonic functions. The explicit form is 
\bea
 c_{\ell_1,m_1;\ell_2,m_2;L,M}=(-1)^M\sqrt{\frac{(2\ell_1+1)(2\ell_2+1)(2L+1)}{4\pi}}\left(\begin{array}{ccc}\ell_1&\ell_2&L\\0&0&0\end{array}\right)\left(\begin{array}{ccc}\ell_1&\ell_2&L\\ m_1&m_2&-M\end{array}\right)\nn
\eea in  terms of Wigner's 3$j$-Symbols.

\item Non-local terms. There are three non-local terms in \eqref{nonmm}, \eqref{nonmq} and \eqref{nonqq}. The non-local terms introduce new operators in the commutators. It is understood that the new operators are also normal ordered. It would be interesting to explore the commutators with these new operators. However, we find an interesting truncation by setting 
\be 
\dot Y=\dot Z=\dot g=\dot{g}_1=\dot{g}_2=0.\label{nnon}
\ee In this case, all the non-local terms and the central terms $C_M,C_Q,C_{TQ}$ are zeros. The reader can find more details in Appendix \ref{commutators}.
\item Truncation \Rmnum{1}. To find the connections to BMS algebra, we impose two conditions on the commutators.
\begin{itemize}
	\item There are only supertranslation generators and superrotation generators in the algebra. The algebra may also include central extension terms which are proportional to the identity operator.
	\item The algebra should be closed and satisfy the Jacobi identity.
\end{itemize}
The truncated algebra is 
\bea 
\    [\mathcal{T}_{f_1},\mathcal{T}_{f_2}]&=&C_T(f_1,f_2)+i\mathcal{T}_{f_1\dot f_2-f_2\dot f_1},\label{vir1}\\
\ [\mathcal{T}_f,\mathcal{M}_Y]&=&-i\mathcal{T}_{Y^A\nabla_Af},\label{vir2}\\
\ [\mathcal{M}_Y,\mathcal{M}_Z]&=&i\mathcal{M}_{[Y,Z]}.\label{vir3}
\eea Since $\mathcal{T}_f$ generates GSTs and $\mathcal{M}_Y$ generates SSRs, we may denote the corresponding group  as
\be 
\text{Diff}(S^2)\ltimes C^\infty(\mathcal{I}^+).\label{virnu}
\ee  The notation $\text{Diff}(S^2)$ means that the vectors $Y^A(\Omega)$ generate diffeomorphisms of $S^2$, and $C^\infty(\mathcal{I}^+)$ means that $f$ is any smooth function on $\mathcal{I}^+$.%This is a generalization of the BMS algebra.
\item Truncation \Rmnum{2}. We can also include the operator $\mathcal{Q}$. To eliminate the non-local terms in the algebra, we still impose the condition \eqref{nnon}. From \eqref{nonlq}, the function $f$ should be a linear function of $u$ 
\be 
f(u,\Omega)=f(\Omega)+u\ h(\Omega).
\ee Now we define two independent operators from $\mathcal{T}_f$
\bea 
\mathcal{P}_f=\int du d\Omega f(\Omega):\dot\Sigma^2:,\quad \mathcal{D}_h=\int du d\Omega \ u\   h(\Omega):\dot\Sigma^2:.
\eea We find the following algebra 
\bea 
\ [\mathcal{P}_{f_1},\mathcal{P}_{f_2}]&=&0,\\
\ [\mathcal{P}_f,\mathcal{D}_h]&=&i\mathcal{P}_{fh},\\
\ [\mathcal{P}_f,\mathcal{M}_Y]&=&-i\mathcal{P}_{Y^A\nabla_A f},\\
\ [\mathcal{P}_f,\mathcal{Q}_g]&=&0,\\
\ [\mathcal{D}_{h_1},\mathcal{D}_{h_2}]&=&0,\\
\ [\mathcal{D}_h,\mathcal{M}_Y]&=&-i\mathcal{D}_{Y^A\nabla_A h},\\
\ [\mathcal{D}_h,\mathcal{Q}_g]&=&i\mathcal{Q}_{hg},\\
\ [\mathcal{M}_Y,\mathcal{M}_Z]&=&i\mathcal{M}_{[Y,Z]},\\
\ [\mathcal{M}_Y,\mathcal{Q}_g]&=&i\mathcal{Q}_{Y^A\nabla_A g},\\ 
\ [\mathcal{Q}_{g_1},\mathcal{Q}_{g_2}]&=&0.
\eea In this algebra, all the functions and vectors are defined on $S^2$ and are independent of $u$. Since there is no geometric meaning for the operator $\mathcal{Q}_g$, we may also truncate it away to find the following subalgebra 
\bea 
\ [\mathcal{P}_{f_1},\mathcal{P}_{f_2}]&=&0,\label{dbms1}\\
\ [\mathcal{P}_f,\mathcal{D}_h]&=&i\mathcal{P}_{fh},\label{dbms2}\\
\ [\mathcal{P}_f,\mathcal{M}_Y]&=&-i\mathcal{P}_{Y^A\nabla_A f},\label{dbms3}\\
\ [\mathcal{D}_{h_1},\mathcal{D}_{h_2}]&=&0,\label{dbms4}\\
\ [\mathcal{D}_h,\mathcal{M}_Y]&=&-i\mathcal{D}_{Y^A\nabla_A h},\label{dbms5}\\
\ [\mathcal{M}_Y,\mathcal{M}_Z]&=&i\mathcal{M}_{[Y,Z]}. \label{dbms6}
\eea This is exactly the algebra found in 
\cite{Donnay:2020fof}.% we will not discuss this algebra in this paper. 
\item BMS algebra. The algebra \eqref{vir1}-\eqref{vir3} can be truncated to the usual BMS algebra. We just list two possible truncations. 
\begin{itemize}
	\item $\mathcal{T}_f$ generates SSTs and $\mathcal{M}_Y$ generates SSRs. Then the central charge becomes zero. The algebra truncates to the BMS algebra in the CL sense. 
	\item $\mathcal{T}_f$ generates SSTs and $\mathcal{M}_Y$ generates global conformal transformations of $S^2$. The algebra truncates to the original BMS algebra.
\end{itemize}
\end{itemize}
\subsection{Antipodal matching condition}\label{antipodalsec}
%This is a non-linear partial differential equation, we may solve it perturbatively by first setting the potential $V(\Phi)=0$.
We can also discuss the symmetry group at past null infinity ($\mathcal{I}^-$). The corresponding radiation phase space could be encoded in the field $\Sigma^-(v,\Omega)$. This field is the leading fall-off term of the field $\Phi$ near $\mathcal{I}^-$
\bea 
\Phi(t,\bm x)=\frac{\Sigma^-(v,\Omega)}{r}+\mathcal{O}\left(\frac{1}{r^2}\right).
\eea 
The coordinate $v=t+r$ is the advanced time in Minkowski spacetime. 
The symmetry groups at $\mathcal{I}^+$ and $\mathcal{I}^-$ may be related to each other by antipodal matching condition \cite{Strominger:2017zoo}. 
In this section, we will derive the antipodal matching condition using two different methods. We will set the potential $V(\Phi)=0$ to simplify the discussion. The equation of motion is a linear partial differential equation 
\be 
\Box\Phi=-J.
\ee 
Using Green's function, the linear equation of motion is solved by \bea 
\Phi(t,\bm x)=\Phi^{\text{in}}(t,\bm x)+\Phi^{\text{ret}}(t,\bm x).
\eea 
The field $\Phi^{\text{in}}(t,\bm x)$ obeys the sourceless Klein-Gordon equation. It is determined by the incoming waves from past null infinity $\mathcal{I}^-$.  The second term is the retarded solution which is caused by the source
\bea 
\Phi^{\text{ret}}(t,\bm x)=\frac{1}{4\pi}\int d\bm x' \frac{J(t-|\bm x-\bm x'|,\bm x')}{|\bm x-\bm x'|}.
\eea 
There is a symmetric solution which is written by 
\bea 
\Phi(t,\bm x)=\Phi^{\text{out}}(t,\bm x)+\Phi^{\text{adv}}(t,\bm x).
\eea Now the field $\Phi^{\text{out}}(t,\bm x)$ is the outgoing waves to $\mathcal{I}^+$. The second term is the advanced solution
\bea 
\Phi^{\text{adv}}(t,\bm x)=\frac{1}{4\pi}\int d\bm x' \frac{J(t+|\bm x-\bm x'|,\bm x')}{|\bm x-\bm x'|}.
\eea The difference between the outgoing waves and the incoming waves
\bea 
\Phi^{\text{rad}}(t,\bm x)=\Phi^{\text{out}}(t,\bm x)-\Phi^{\text{in}}(t,\bm x)=\Phi^{\text{ret}}(t,\bm x)-\Phi^{\text{adv}}(t,\bm x)\label{radiationfield}
\eea  may be regarded as the radiation field  \cite{Dirac:1938nz}. Near $\mathcal{I}^+$, the advanced solution is zero. 
Using the Fourier transformation of the source 
\bea
J(t,\bm x)&=&\int \frac{d\omega d\bm k}{(2\pi)^4}J(\omega,\bm k)e^{-i\omega t+i\bm k\cdot\bm x},
\eea we can find the large $r$ expansion of the field $\Phi^{\text{rad}}$  and read out the leading term
\bea 
\Sigma(u,\Omega)=\frac{1}{8\pi^2}\int_{-\infty}^\infty d\omega J(\omega,\bm k)e^{-i\omega u},\quad \bm k=(\omega,\theta,\phi).\label{futuresigma}
\eea 
Usually, the source is located at a finite region of space. It will contribute to the classical solution of the field $\Sigma$. For example, for a point source whose frequency is $\omega_0$ and location is the origin, 
\be 
J(t,\bm x)=\cos\omega_0 t\  \delta(\bm x),
\ee we find the radiation field 
\bea 
\Sigma(u,\Omega)=\frac{\cos\omega u}{4\pi}.
\eea 
Similarly, near $\mathcal{I}^-$, only the advanced solution in \eqref{radiationfield} is relevant. We find 
\bea
\Sigma^-(v,\Omega)=-\frac{1}{8\pi^2}\int_{-\infty}^{\infty} d\omega e^{-i\omega v}J(\omega,-\bm k).\label{pastsigma}
\eea In spherical coordinates, 
\be 
\bm k=(\omega,\theta,\phi)=(\omega,\Omega),\quad -\bm k=(\omega,\pi-\theta,\pi+\phi)=(\omega,\Omega^P),
\ee where $\Omega^P$ is the antipodal point of $\Omega=(\theta,\phi)$ 
\bea 
\Omega^P=(\pi-\theta,\pi+\phi).
\eea 
The antipodal point of $\Omega^P$ is still $\Omega$
\be 
(\Omega^P)^P=\Omega. 
\ee 
Comparing the solution \eqref{pastsigma} with \eqref{futuresigma}, we find 
\bea
J(\omega,\bm k)=\frac{1}{4\pi}\int du e^{i\omega u} \Sigma(u,\Omega)=-\frac{1}{4\pi}\int dv e^{i\omega v}\Sigma^{-}(v,\Omega^P).
\eea We may define the Fourier transformation 
\bea 
\Sigma(u,\Omega)&=&\frac{1}{2\pi}\int_{-\infty}^\infty d\omega e^{-i\omega u} \Sigma(\omega,\Omega),\\
\Sigma^-(v,\Omega)&=&\frac{1}{2\pi}\int_{-\infty}^\infty d\omega e^{-i\omega v} \Sigma^-(\omega,\Omega).
\eea 
Then we find the following antipodal identification
\bea
\Sigma(\omega,\Omega)=-\Sigma^{-}(\omega,\Omega^P). \label{antipodalcondition}
\eea %In the soft limit $\omega\to 0$, the antipodal matching condition becomes
%\bea 
%\Sigma(0,\Omega)=-\Sigma^-(0,\Omega^P).
%\eea 
%This condition can also be expressed as 
%\bea 
%\Sigma^-(v,\Omega)=-\Sigma(u,\Omega')|_{u\to v,\Omega'\to -\Omega}.\label{matching}
%\eea 
The antipodal matching condition \eqref{antipodalcondition} can also be proved at the quantum level. Following the same procedure as subsection \ref{comuscip}, we find the mode expansion of the field $\Sigma^-(v,\Omega)$ at $\mathcal{I}^-$
 \bea 
\Sigma^-(v,\Omega)&=&\int_0^\infty \frac{d\omega}{\sqrt{4\pi\omega}}\sum_{\ell m}[\bar{a}_{\omega,\ell,m}e^{-i\omega v}Y_{\ell,m}(\Omega)+\bar{a}^\dagger_{\omega,\ell,m}e^{i\omega v}Y^*_{\ell,m}(\Omega)]\label{sigmaexpscim}
\eea
 where
\bea
\bar{a}_{\omega,\ell,m}&=&(-1)^{\ell}\frac{\omega i }{2\sqrt{2}\pi^{3/2} }\int d\Omega b_{\bm k}Y_{\ell,m}^*(\Omega),\label{annihiscrim}\\
\bar{a}_{\omega,\ell,m}^\dagger&=&(-1)^\ell \frac{\omega }{2\sqrt{2}\pi^{3/2}i}\int d\Omega b^\dagger_{\bm k}Y_{\ell,m}(\Omega).\label{createscrim}
\eea Comparing them with the equation \eqref{annihi} and \eqref{create}, we find the following matching condition 
\bea 
\bar{a}_{\omega,\ell,m}=(-1)^{\ell+1}a_{\omega,\ell,m},\hspace{1cm} \bar{a}_{\omega,\ell,m}^\dagger=(-1)^{\ell+1}a^\dagger_{\omega,\ell,m}.\label{matchinga}
\eea 
Using the inverse Fourier transformation 
\bea 
\Sigma(\omega,\Omega)&=&\int_{-\infty}^\infty du e^{i\omega u}\Sigma(u,\Omega),\\
\Sigma^-(\omega,\Omega)&=&\int_{-\infty}^\infty dv e^{i\omega v}\Sigma^-(v,\Omega), 
\eea we find 
\bea 
\Sigma(\omega,\Omega)&=&\theta(\omega)\sqrt{\frac{\pi}{\omega}}\sum_{\ell,m}a_{\omega,\ell,m}Y_{\ell,m}(\Omega)+\theta(-\omega)\sqrt{-\frac{\pi}{\omega}}\sum_{\ell,m}a_{-\omega,\ell,m}^\dagger Y^*_{\ell,m}(\Omega),\\
\Sigma^-(\omega,\Omega)&=&\theta(\omega)\sqrt{\frac{\pi}{\omega}}\sum_{\ell,m}\bar{a}_{\omega,\ell,m}Y_{\ell,m}(\Omega)+\theta(-\omega)\sqrt{-\frac{\pi}{\omega}} \sum_{\ell,m}\bar{a}_{-\omega,\ell,m}^\dagger Y^*_{\ell,m}(\Omega)
\eea Substituting the matching condition for the creation and annihilation operators \eqref{matchinga} and using the 
 parity transformation of the spherical harmonic functions
\be 
Y_{\ell,m}(\Omega)=(-1)^\ell Y_{\ell,m}(\Omega^P),
\ee 
we find 
\bea 
\Sigma(\omega,\Omega)=-\Sigma^-(\omega,\Omega^P).
\eea This is the same matching condition as \eqref{antipodalcondition}.

%\section{Virasoro algebra in higher dimensions}
%\section{Superrotation}
\section{Geometric approach}\label{nusec}
The BMS group could be regarded as a geometric symmetry of the Carroll manifold $\mathcal{I}^+$. 
Following \cite{Duval_2014a,Duval_2014b,Duval:2014uoa}, we will review the conformal Carroll group and the Newman-Unti group in this section. It turns out that the symmetry group we found in the previous section is an extension of the Newman-Unti group. %In this section, we will treat the BMS group as a conformal symmetry of the Carroll manifold $\mathcal{I}^+$. The symmetry group we found in previous sections is an extension of the Newman-Unti group.
\subsection{Conformal Carroll group}
The future null infinity $\mathcal{I}^+$ is a Carroll manifold 
\be 
\mathcal{I}^+=\mathbb{R}\times S^2.
\ee This is a null hypersurface with a singular metric 
\be 
ds^2=\gamma_{AB}d\theta^Ad\theta^B.
\ee To generate the retarded time direction, one should introduce a vector which is the kernel of the metric $\gamma_{AB}$
\be 
\chi=\partial_u.
\ee 
The conformal Carroll group of level $k$ 
\be \text{CCarr}_k(\mathcal{I}^+,\gamma,\chi)
\ee  is generated by the vector $\xi$ such that 
\bea 
\mathcal{L}_\xi \gamma=\lambda \gamma,\quad \mathcal{L}_\xi \chi=\mu\chi,\quad \lambda+k \mu=0,
\eea where $\lambda$ and $\mu$ are conformal factors. Then the vector $\xi$ is 
\bea 
\xi=Y^A(\Omega)\partial_A+(f(\Omega)+\frac{u}{k}\nabla_A Y^A(\Omega))\partial_u.\label{ccglk}
\eea The vector $Y^A(\Omega)$ is a Conformal Killing vector of the sphere 
\be 
\nabla_A Y_B+\nabla_B Y_A=\gamma_{AB}\nabla_C Y^C.\label{ckvY}
\ee The conformal Carroll group of level $k$ is a semi-direct product of conformal transformations of $S^2$ and SSTs.
%\be 
%\text{CCarr}_k(\mathcal{I}^+,\gamma,\chi)=\text{Conf}(S^2)\ltimes C^\infty({S}^2).
%\ee 
When $k=2$, the algebra  \eqref{ccglk} is exactly the standard BMS algebra.
\subsection{Newman-Unti group}
Newman-Unti group 
\be 
\text{NU}(\mathcal{I}^+,\gamma,\chi)
\ee  is one of the extensions of the conformal Carroll group. It is generated by the vectors $\xi$  that preserve the conformal structure of the metric $\gamma$ 
\be 
\mathcal{L}_\xi \gamma=\lambda \gamma.\label{confgamma}
\ee In this case, the vector $\xi$ is 
\be 
\xi=Y^A(\Omega)\partial_A+f(u,\Omega)\partial_u,  \label{nuxi}
\ee 
where the vector $Y^A$ is still a conformal Killing vector of $S^2$. However, the function $f$ may depend on the retarded time $u$. Therefore, the Newman-Unti group is a semi-direct product of conformal transformations of $S^2$ and GSTs
\be 
\text{NU}(\mathcal{I}^+,\gamma,\chi)=\text{Conf}(S^2)\ltimes C^\infty(\mathcal{I}^+).\label{nuconformal}
\ee It is possible to truncate it to the Newman-Unti group of level $k\ (k=1,2,3,\cdots)$ by requiring 
\be 
\mathcal{L}_\xi \gamma=\lambda \gamma,\quad (\mathcal{L}_\chi)^k\xi=0.
\ee We still find the vector \eqref{nuxi}. Besides the constraint \eqref{ckvY}, the GST should be a  polynomial of $u$ with degree $k-1$ 
\bea 
f(u,\Omega)=\sum_{n=1}^{k} u^{n-1}f_n(\Omega).
\eea 
\subsection{Extension of the Newman-Unti group}\label{sec:Newman-Unti}
In two-dimensional Virasoro algebra, the central charge is related to conformal anomaly. When the central charge is zero, the Virasoro algebra becomes the Witt algebra. Similarly, We may set the central charge $C_T(f_1,f_2)$ to be zero and find the classical version of the algebra \eqref{vir1}-
\eqref{vir3}
\bea 
[\mathcal{T}_{f_1},\mathcal{T}_{f_2}]&=&i\mathcal{T}_{f_1\dot f_2-f_2\dot f_1},\label{c1}\\
\ [\mathcal{T}_f,\mathcal{M}_Y]&=&-i\mathcal{T}_{Y^A\nabla_Af},\label{c2}\\
\ [\mathcal{M}_Y,\mathcal{M}_Z]&=&i\mathcal{M}_{[Y,Z]}.\label{c3}
\eea Interestingly, this algebra is realized by the vector 
\be 
\xi=f(u,\Omega)\partial_u+Y^A(\Omega)\partial_A.\label{genbms}
\ee It is easy to see that the group \eqref{virnu} is a straightforward generalization of the Newman-Unti group \eqref{nuconformal}. %In section \ref{reviewform}, we have mentioned the Newman-Unti group for the Carroll manifold $\mathcal{I}^+$ with metric $\gamma$ and vector $\chi$. The Newman-Unti group is generated by all transformations that preserve the conformal structure of the sphere 
%\be 
%\mathcal{L}_\xi \gamma=\lambda \gamma.\label{confgamma}
%\ee 
To generalize the Newman-Unti group, we should  abandon the condition 
\eqref{confgamma} and impose the condition
\be 
\mathcal{L}_\xi \chi=\mu\chi.\label{confchi}
\ee The most general solution of \eqref{confchi} is exactly \eqref{genbms}. The function $\mu$ is 
\be 
\mu=-\dot f(u,\Omega).
\ee The  Lie derivative of the metric along \eqref{genbms} is still singular
\bea 
\mathcal{L}_\xi \gamma_{uu}=\mathcal{L}_\xi \gamma_{uA}=0,\quad \mathcal{L}_\xi \gamma_{AB}=\nabla_A Y_B(\Omega)+\nabla_BY_A(\Omega).
\eea Therefore, the vector $\chi$ is still the kernel of the metric after the transformation. Now we consider the vector with respect to GSTs and GSRs
\be 
\widetilde{\xi}=f(u,\Omega)\partial_u+Y^A(u,\Omega)\partial_A.\label{gstgsr}
\ee We find 
\bea 
&&\mathcal{L}_{\widetilde{\xi}}\chi^u=-\dot f(u,\Omega),\quad \mathcal{L}_{\widetilde{\xi}}\chi^A=-\dot Y^A(u,\Omega),\\
&& \mathcal{L}_{\widetilde{\xi}}\gamma_{uu}=0,\quad \mathcal{L}_{\widetilde{\xi}}\gamma_{uA}=\dot{Y}_A(u,\Omega),\quad \mathcal{L}_{\widetilde{\xi}}\gamma_{AB}=\nabla_AY_B(u,\Omega)+\nabla_BY_A(u,\Omega).
\eea The manifold is not Carrollian after the transformation \eqref{gstgsr}. The GSRs break the null structure of $\mathcal{I}^+$. This may interpret why we should consider GSTs and SSRs to form a closed algebra. Interestingly, the finite transformation of \eqref{genbms} is exactly the Carrollian diffeomorphism defined in \cite{Ciambelli:2018xat,Ciambelli:2019lap}. From a geometric point of view, we may define any consistent field theory on $\mathcal{I}^+$. When there is no anomaly, they should obey the geometric symmetry \eqref{c1}-\eqref{c3}. In other words, there should be corresponding generators with respect to the vector \eqref{genbms}. They are exactly the supertranslation and superrotation generators.

We will further comment on the structure \eqref{dbms1}-\eqref{dbms6}. This algebra is generated by the vector 
\bea 
\xi=[f(\Omega)+u\  h(\Omega)]\partial_u+Y^A(\Omega)\partial_A.\label{dbms}
\eea 
This is obtained by reducing the function $f(u,\Omega)$  in the generator \eqref{genbms} to a linear polynomial of $u$. Similar to the Newman-Unti group of level $k$, we may define its extension as 
\bea 
\mathcal{L}_\xi \chi=\mu\chi,\quad (\mathcal{L}_\chi)^k \xi=0.
\eea By setting $k=2$, the solution is exactly \eqref{dbms}.

%\textbf{Newmann Unti}
\section{Conclusion and discussion}\label{dissec}
In this paper, we reduce the massless scalar field theory in Minkowski spacetime to future null infinity $\mathcal{I}^+$. The information of the scalar field is encoded in a single field $\Sigma$ at $\mathcal{I}^+$.  The ten Poincar\'e fluxes are totally determined by the field $\Sigma$. We obtain the flux operators and interpret them as supertranslation and superrotation generators. These flux operators do not form a closed algebra in general. However, there is a consistent group which is formed by GSTs and SSRs. Its classical version, as a generalization of the Newman-Unti group, could be realized as a geometric symmetry of the Carroll manifold $\mathcal{I}^+$. 

We notice that the subalgebra \eqref{vir1} is a Virasoro algebra. This has been found in the context of light-ray operator \cite{Korchemsky:2021htm}. Other works on Virasoro algebra from light ray operator include \cite{Huang:2019fog,Huang:2020ycs,Huang:2021hye,Belin:2020lsr,Besken:2020snx}.
However, there are subtle differences between their results and ours. In \cite{Korchemsky:2021htm}, the four-dimensional Virasoro algebra is realized by free fermion or Maxwell theory. For free scalar theory, the authors found a non-local term in the commutator of their energy flow operators, see equation (1.9) of \cite{Korchemsky:2021htm}. Switching into our language, the energy flow operator defined in their paper corresponds to the operator 
	\be 
	\widetilde{\mathcal{T}}_f=\mathcal{T}_f-\frac{1}{6}\mathcal{Q}_{\ddot f}.\label{wdtt}
	\ee We have checked that the commutator $[\widetilde{\mathcal{T}}_{f_1},\widetilde{\mathcal{T}}_{f_2}]$  (more precisely $[\widetilde{\mathcal{T}}_{\omega,\ell,m},\widetilde{\mathcal{T}}_{\omega',\ell',m'}]$) is  exactly equivalent to equation (1.9) of \cite{Korchemsky:2021htm}. See Appendix \ref{Light-ray} for more details. Therefore, there is no contradiction with light ray algebra. The algebra \eqref{vir1}-\eqref{vir3} could be regarded as a direct generalization of Virasoro algebra with superrotation. 
It has been known for several years that the BMS algebra could be realized as a light ray algebra \cite{Cordova:2018ygx}. However, the work of \cite{Cordova:2018ygx} only focused on average null operators. They correspond to the soft limit in the Fourier space.  Our result could also be regarded as an extension of \cite{Cordova:2018ygx} away from the soft limit.

There are various open questions in this direction. 
\begin{itemize}
	\item More general field theories. We mainly focus on massless free scalar theory in this work. However, we could explain the group we found as a geometric symmetry of $\mathcal{I}^+$. This implies that the algebra \eqref{vir1}-\eqref{vir3} may be valid for much more field theory. For Maxwell theory and gravitational theory, we may check this point. Since the number of the propagating degrees of freedom is 2 for these theories, the central charge is two times with respect to the real scalar theory. For interacting field theory, it is interesting to see whether it is possible refine the energy flow operators defined in \cite{Korchemsky:2021htm} such that the algebra \eqref{vir1}-\eqref{vir3} is preserved. 
	\item Field theories on the Carroll manifold $\mathcal{I}^+$. We notice that the two-point correlator of the scalar field in  \eqref{cordotsigdotsig} matches with \cite{Chen:2021xkw} from representation theory of Carrollian conformal field theory. By dimension analysis, we find $[\Phi]=1$, $[\Sigma]=0$ and $[\dot\Sigma]=1$. It follows that correlation function $\braket{0|\dot\Sigma\dot\Sigma|0}$ is expected to be proportional to $(u-u')^{-2}$, just as in \cite{Chen:2021xkw}. There are also Carrollian free scalar models in the literature \cite{Bagchi:2019xfx,Henneaux:2018mgn,Bagchi:2019clu,Hao:2021urq,Bagchi:2022emh,Bagchi:2022eav}. However, we should emphasize that our results are not based on the existence of any action on $\mathcal{I}^+$. As a consequence, there is no equation of motion for the boundary field $\Sigma$. The solution phase space is larger than the Carrollian free scalar model. Consequently, we find a much larger group than BMS group. Actually, the symmetry group can be extended further by including higher spin fluxes \cite{Campoleoni:2017mbt,Campoleoni:2021blr, Bekaert:2022ipg}. Since the symmetry  \eqref{vir1}-\eqref{vir3} could be understood as a geometric symmetry of the Carroll manifold $\mathcal{I}^+$ classically, one may also consider representation theory of this symmetry group and define field theory on $\mathcal{I}^+$.  The field theory on $\mathcal{I}^+$ may provide explicit realization of flat holography. 
 \item Non-local terms. As we have discussed in section 6, for the Carrollian diffeomorphism which is generated by GSTs and SSRs, there is no non-local term in the algebra. They appear only for GSRs. As we have shown in \eqref{GSRtransint}, the non-local term is the obstacle to  identifying $\mathcal{M}_Y$ as a superrotation generator for the case of GSRs. As we expect, GSRs violate the null structure of the Carrollian manifold. The violation may be reflected in the non-local terms and may be thought as their origin. It is interesting to discuss this topic in the future.
 
	\item Correlators. In the context of conformal collider physics \cite{Hofman:2008ar}, the energy correlators correspond to the correlators of the soft limit of the supertranslation generators. From the point of view of BMS algebra, it is also interesting to consider the correlators of the superrotation generators. They may relate to angular momentum correlators. 
	\item Regularization of the central charge. The central charge is divergent in our Virasoro algebra. It would be fine to find a way to regularize it.  To find the physical meaning of this central charge, we use the completeness of the spherical harmonic function 
\bea 
\delta^{(2)}(0)&=&\sum_{\ell,m}Y_{\ell,m}(\Omega)Y_{\ell,m}^*(\Omega)\nn\\&=&\frac{1}{4\pi}\sum_{\ell=0}^\infty (2\ell+1)P_{\ell}(0)\nn\\&=&\frac{1}{4\pi}\sum_{\ell,m}1.
\eea At the second line, we used the addition theorem of spherical Harmonic function. The Legendre function $P_\ell(x)$ has the special value 
\be 
P_\ell(0)=1.
\ee At the last step, we used the fact that there are $2\ell+1$ spherical harmonic functions for each $\ell$.   Interestingly,  it is clear that 
\bea 
\delta^{(2)}(0)&=&\frac{\text{Number of independent states on the unit $S^2$}}{\text{Area of the unit $S^2$}}\nn\\&=&\text{Density of states on the unit $S^2$}.
\eea Therefore, roughly speaking,  $\delta^{(2)}(0)$ counts the degrees of freedom on the sphere. In two-dimensional conformal field theory, the central charge also counts the  degrees of freedom of the theory. Unfortunately, the number of the  degrees of freedom on the sphere is infinity. One should find a way to regularize the Dirac delta function. A naive method is to use zeta function regularization 
\bea 
\delta^{(2)}(0)=\frac{1}{4\pi}\sum_{\ell=0}^\infty (2\ell+1)=\frac{1}{4\pi}[1+2\zeta(-1)+\zeta(0)]=\frac{1}{12\pi}.
\eea We will not discuss more on the regularization of Dirac delta function on the sphere. It would be interesting to check this regularization method in the future. 
\item  Relaxed fall-off conditions. As we have emphasized, the GSTs and GSRs are defined through the Fourier transformation of the  energy and angular momentum  flux density operators in this work. Strictly speaking, they may violate the usual fall-off conditions in the context of BMS symmetry. It is natural to see whether one can relax fall-off conditions and explore the BMS group further up. 
%\textbf{light ray}
\end{itemize}
\vspace{10pt}
{\noindent \bf Acknowledgments.} %J.L. thanks for the organisers of ``the Third National Conference on Qauntum Field Theory and String Theory''.
The work of J.L. is supported by NSFC Grant No. 12005069.
\appendix

\section{Conformal Killing vectors on $S^2$}
\label{ckvs}
The metric for a unit sphere $S^2$ is 
\begin{align}
  ds^2=\gamma_{AB}d\theta^Ad\theta^B,\qquad A,B=1,2.
\end{align}
The conformal Killing vector (CKV) is the vector $Y^A$ that obeys the equation 
\begin{align}
  \nabla_AY_B+\nabla_BY_A=\gamma_{AB}\nabla_CY^C.
\end{align}
There are six global solutions for this equation. 
\begin{itemize}
	\item There are three Killing vectors on $S^2$, which are denoted as $Y^A_{ij}$ in the context. The subscript $ij$ are antisymmetric
  \begin{align}
    Y^A_{ij}=-Y^A_{ji},\qquad i,j=1,2,3.
  \end{align}
  They satisfy the following condition 
  \begin{align}
    \nabla_A Y^A_{ij}=0.
  \end{align}
  \item There are three strictly conformal Killing vectors on $S^2$, denoted as $Y^A_i$ in this paper. The subscript $i=1,2,3$. Their divergences are not zero but \(2n_i\)
  \begin{align}
    \nabla_AY^A_i=2n_i.
  \end{align}
\end{itemize}
These six CKVs generate the group $SO(1,3)$, satisfying the commutation relations below
\begin{align}
  & [Y_i, Y_j]=Y_{ij}, \\
  &  [Y_{ij}, Y_k]=\delta_{jk}Y_i-\delta_{ik}Y_j, \\
  & [Y_{ij}, Y_{kl}]=-\delta_{ik}Y_{jl}+\delta_{jk}Y_{il}-\delta_{jl}Y_{ik}+\delta_{il}Y_{jk}.
\end{align}
We collect some useful identities in the following.
\begin{enumerate}
  \item Killing vectors and strictly conformal Killing vectors are related by
  \begin{align}
    Y^A_{ij}=Y^A_in_j-Y^A_j n_i.
  \end{align}
  The reverse relation is
  \begin{align}
    Y^A_i=Y_{ij}^An_j.
  \end{align}
  \item Identities involving the products of normal vectors and CKVs are
  \begin{align}
    n_iY_i^A=0,\qquad \epsilon_{ijk}n_iY_{jk}^A=0.
  \end{align}
  \item Derivatives of normal vector are
  \begin{align}
    \nabla^An_i=-Y_i^A,\qquad \partial_in_i=\frac{2}{r}.
  \end{align}
  \item It is easy to find that
  \begin{align}
    Y^A_i Y^B_j \delta_{ij}=\gamma^{AB},\qquad Y^{ij\ A}Y_{ij}^B=2\gamma^{AB}.
  \end{align} 
  We also have identities with angular indices contracted
  \begin{align}
    &Y_{ij}\cdot Y_k\equiv Y_{ij}^A \gamma_{AB}Y^B_k=\delta_{ik}n_j-\delta_{jk}n_i,\\
    &Y_i\cdot Y_j+n_in_j\equiv Y_i^A\gamma_{AB} Y_j^B+n_in_j=\delta_{ij},
  \end{align}
  Other identities related to products of CKVs are collected below
  \begin{align}
    &Y_{ij}^AY^B_j+Y_{ij}^BY^A_j=-2\gamma^{AB}n_i,\\
    &Y_i^AY_j^B-Y_i^BY_j^A=\epsilon^{AB}\epsilon_{ijk}n^k.
  \end{align}
\end{enumerate}

\section{ Symplectic structure}\label{hamiltonmethod}
The action of scalar field is
\begin{align}
  S=\int d^4x\sqrt{-g}\ [-\frac{1}{2}\partial_\mu\Phi\partial^\mu\Phi -V(\Phi)].
\end{align}
Its variation under $\delta\Phi$ reads as
\begin{align}
  S%=&-\int\delta(\partial_\mu\Phi)\partial^\mu\Phi \sqrt{-g} d^4x\nn\\
  =&\int d^4x\sqrt{-g}\big[-\partial_\mu(\partial^\mu\Phi\delta\Phi)+(\partial_\mu\partial^\mu\Phi-V'(\Phi))\delta\Phi\big].
\end{align}
It is easy to see that the second term is proportional to the bulk equation of motion. We could write out the presymplectic potential form
\begin{align}
  \pmb \Theta(\delta\Phi;\Phi)=-\partial^\mu\Phi\delta\Phi (d^3x)_\mu,
\end{align}
where
\begin{align}
  (d^3x)_\mu=\frac{1}{6}\epsilon_{\mu\nu\rho\sigma}dx^\nu\wedge dx^\rho\wedge dx^\sigma.
\end{align}
Therefore, we obtain the presymplectic form
\begin{align}
  \pmb \omega(\delta_1\Phi,\delta_2\Phi;\Phi)=\delta_1\pmb \Theta(\delta_2\Phi;\Phi)-\delta_2\pmb \Theta(\delta_1\Phi;\Phi)=-(d^3x)_\mu[\delta_1(\partial^\mu\Phi)\delta_2\Phi-\delta_2(\partial^\mu\Phi)\delta_1\Phi] .
\end{align}

%Using the fall-off condition \eqref{bc}, one can work out
%\begin{align}
 % \delta_2(\partial^r\Phi)\delta_1\Phi=-r^{-2}\delta_2\dot\Sigma\delta_1\Sigma+{\cal O}(r^{-3}).
%\end{align}
Given the fall-off condition \eqref{bc},  the presymplectic form  at null boundary  becomes
\begin{align}
  \pmb \omega(\delta_1\Phi,\delta_2\Phi;\Phi)=-\sin\theta du\wedge d\theta\wedge d\phi[\delta_1\dot\Sigma\delta_2\Sigma-\delta_2\dot\Sigma\delta_1\Sigma]+{\cal O}(r^{-1}).
\end{align}
Eventually, one could obtain the commutator of the propagating field $\Sigma$
\begin{align}
  [\Sigma(u,\Omega),\dot\Sigma(u',\Omega')]=\frac{i}{2}\delta(u-u')\delta(\Omega-\Omega').
\end{align}
With this commutator at hand, other commutation relations and the correlation functions of $\Sigma$ are easy to derive.

\section{Commutators}
\label{commutators}
In this appendix, we will discuss the computation of the commutators.
\begin{itemize}
	\item Central charges. The central charge term can be found from the two-point correlators. As an example, we compute the central charge term in the commutator $[\mathcal{T}_{f_1},\mathcal{T}_{f_2}]$. We first note that the supertranslation generator is constructed from the local operator $T(u,\Omega)$. The two-point correlators can be found in \eqref{TT2pt}. When two operators $T(u,\Omega)$ and $ T(u',\Omega')$ are close enough, we can write the operator product expansion schematically as 
	\be 
	 TT\sim \bm 1+\cdots.
	\ee On the right hand side, we just write down the term which is proportional to the identity. The terms in $\cdots$ are vanishing when we compute the correlator $\langle 0| [T, T]|0\rangle$. Therefore, to find the central extension, it is enough to know the two-point correlator
	\bea 
	\ && \text{Central term of } [\mathcal{T}_{f_1},\mathcal{T}_{f_2}]=\langle 0|[\mathcal{T}_{f_1},\mathcal{T}_{f_2}]|0\rangle\nn\\&=&\int_{\text{Im}(u-u')<0} du d\Omega  du'd\Omega' f_1(u,\Omega)f_2(u',\Omega')\langle 0|T(u,\Omega)T(u',\Omega')|0\rangle\nn\\&&-\int_{\text{Im}(u-u')>0} du d\Omega  du'd\Omega' f_1(u,\Omega)f_2(u',\Omega')\langle 0|T(u',\Omega')T(u,\Omega)|0\rangle\nn\\&=&\int du d\Omega du'd\Omega' f_1(u,\Omega)f_2(u',\Omega')\frac{c}{8\pi^2}\left[\frac{1}{(u-u'-i\epsilon)^4}-\frac{1}{(u'-u-i\epsilon)^4}\right]\delta(\Omega-\Omega')\nn\\&=&-\int du d\Omega du' f_1(u,\Omega)f_2(u',\Omega)\frac{c}{8\pi^2}\frac{1}{6}\partial_u^3\left[\frac{1}{u-u'-i\epsilon}+\frac{1}{u'-u-i\epsilon}\right]\nn\\&=&-\frac{i c}{24\pi}\int du d\Omega du' f_1(u,\Omega)f_2(u',\Omega)\partial_u^3 \delta(u-u')\nn\\&=&-\frac{ic}{48\pi}\int du d\Omega (f_1\dddot {f_2}-f_2\dddot {f_1}).
	\eea 
	\item Non-local terms. We will compute the non-local terms in \eqref{nonmq}. The non-local terms in this commutator are from the non-local term in the commutator $[\mathcal{M}_Y,\Sigma]$
	\bea 
&&	\text{Non-local term of} \  [\mathcal{M}_Y,\mathcal{Q}_g]\nn\\&=&	\text{Non-local term of} \int du d\Omega g(u,\Omega) [\mathcal{M}_Y,:\Sigma^2(u,\Omega):]\nn\\&=&\text{Non-local term of} \ 2\int du d\Omega g(u,\Omega) :\Sigma(u,\Omega) [\mathcal{M}_Y,\Sigma(u,\Omega)]:\nn\\&=&i\int du d\Omega g(u,\Omega):\Sigma(u,\Omega) \int du' \alpha(u'-u) \Delta(\dot{Y};\Sigma;u',\Omega):\nn\\&=&i\int du du'd\Omega \alpha(u'-u)g(u,\Omega):\Sigma(u,\Omega)\Delta(\dot Y;\Sigma;u',\Omega):.
	\eea  

\item The commutator $[{\cal M}_Y,{\cal M}_Z]$. Using the factor $\Delta(Y;\Sigma;u,\Omega)$, we can rewrite the superrotation generator as
\begin{align}
  \mathcal{M}_Y=\int dud\Omega\dot\Sigma\Delta(Y;\Sigma;u,\Omega).
\end{align}
With this expression, we could calculate the commutator $[{\cal M}_Y,{\cal M}_Z]$ as follows. Note that the central charge term could be read out from correlation function using the previous method, and the following calculation only involves terms with fields. Hence, all the terms are not expressed in normal order.
\begin{align}
 \ [\mathcal{M}_Y,\mathcal{M}_Z]= &\int du'd\Omega'[\mathcal{M}_Y,\dot\Sigma(u',\Omega')\Delta(Z;\Sigma;u',\Omega')]\nn\\
  =&i\int du'd\Omega'[\Delta(Z; \dot\Sigma)\Delta(Y; \Sigma)-\Delta(Z; \Sigma)\Delta(Y; \dot\Sigma)]\nn\\
  &+\frac{i}{2}\int du du'd\Omega' \alpha(u-u')\Delta(\dot{Y}; \Sigma; u, \Omega')\Delta(\dot{Z}; \Sigma; u', \Omega').
\end{align}
For the local terms, we find 
\begin{align}
    \Delta(Z; \dot\Sigma)\Delta(Y; \Sigma)-\Delta(Z; \Sigma)\Delta(Y; \dot\Sigma)=\dot\Sigma\big[\Delta(Y;\Delta(Z;\Sigma))-\Delta(Z;\Delta(Y;\Sigma))\big].
\end{align}
There is an identity
\begin{align}
    \Delta(Y;\Delta(Z;\Sigma))-\Delta(Z;\Delta(Y;\Sigma))=\Delta([Y,Z];\Sigma),
\end{align}
with which we can concisely express the local terms as $i{\cal M}_{[Y,Z]}$. 
To prove this identity, we calculate straightforwardly
\begin{align}
  &\Delta(Y;\Delta(Z;\Sigma))-\Delta(Z;\Delta(Y;\Sigma))\nn\\
  =&\frac{1}{2}\nabla_AY^A(Z^B\nabla_B\Sigma+{\frac{1}{2}\nabla_BZ^B\Sigma})+Y^A\nabla_A(Z^B\nabla_B\Sigma+\frac{1}{2}\nabla_BZ^B\Sigma)\nn\\
  &-\frac{1}{2}\nabla_AZ^A(Y^B\nabla_B\Sigma+{\frac{1}{2}\nabla_BY^B\Sigma})-Z^A\nabla_A(Y^B\nabla_B\Sigma+\frac{1}{2}\nabla_BY^B\Sigma)\nn\\
  =&(Y^B\nabla_BZ^A-Z^B\nabla_BY^A)\nabla_A\Sigma+\frac{1}{2}\nabla_A(Y^B\nabla_BZ^A-Z^B\nabla_BY^A)\Sigma\nn\\
  =&\ \Delta([Y,Z];\Sigma).
\end{align}
To obtain the second equality, we have used the following identity
\[Y^A\nabla_A\nabla_B Z^B-Z^B\nabla_B\nabla_AY^A=\nabla_A[Y,Z]^A,\]
for smooth vector on sphere. It follows from the definition of Riemann tensor on sphere
\begin{align}
    [\nabla_A,\nabla_B]V^C=R^C{}_{DAB}V^D.
\end{align}
Riemann tensor on the sphere is 
\be 
R_{ABCD}=\gamma_{AC}\gamma_{BD}-\gamma_{AD}\gamma_{BC}.
\ee 
\end{itemize} 

When the condition \eqref{nnon} is satisfied, we will prove the following two statements.

\begin{itemize} 
	\item The central charges are zeros
\be 
C_M=C_{TQ}=C_{Q}=0.
\ee 
The vanishing of $C_{TQ}$ can be found by setting $\dot g=0$. Therefore, we just need to compute $C_M$ and $C_Q$. We consider the central charge $C_Q$ firstly,
\bea 
C_Q(g_1,g_2)&=&2c \int du du' d\Omega g_1(\Omega)g_2(\Omega)[\beta(u-u')-\beta(u'-u)]\times [\beta(u-u')+\beta(u'-u)]\nn\\&=&ic \int du du'd\Omega g_1(\Omega)g_2(\Omega)\alpha(u-u')\times [\beta(u-u')+\beta(u'-u)]\nn\\&=&0.
\eea At the second line, we used the identity \eqref{alphabeta}. Since $\alpha(u-u')$ is antisymmetric while $\beta(u-u')+\beta(u'-u)$ is symmetric, the integral is exactly zero. Now we compute the central charge 
\bea 
C_M(Y,Z)&=&\widetilde{c}\int du du' \eta(u-u').
\eea We have defined the constant
\be 
\widetilde{c}=\int d\Omega d\Omega' Y^A(\Omega)Z^{B'}(\Omega')\Lambda_{AB'}(\Omega,\Omega')
\ee The function $\eta(u-u')$ can be written as 
\bea 
\eta(u-u')&=&\left[\beta(u-u')-\frac{1}{4\pi}\right]\frac{1}{8\pi}\partial_u \frac{1}{u-u'-i\epsilon}-\left[\beta(u'-u)-\frac{1}{4\pi}\right]\frac{1}{8\pi}\partial_{u'} \frac{1}{u'-u-i\epsilon}\nn\\&=&\partial_u (\cdots)+\partial_{u'}(\cdots)+\frac{1}{32\pi^2(u-u'-i\epsilon)^2}-\frac{1}{32\pi^2(u'-u-i\epsilon)^2}\nn\\&=&\partial_u(\cdots)+\partial_{u'}(\cdots)-\frac{1}{32\pi^2}\partial_u\left[ \frac{1}{u-u'-i\epsilon}+\frac{1}{u'-u-i\epsilon}\right]\nn\\&=&\partial_u(\cdots)+\partial_{u'}(\cdots)-\frac{i}{16\pi}\delta'(u-u').
\eea The notation $\partial_u(\cdots)$ means the corresponding term is a surface term. They do not contribute to the central charge. Therefore,
\bea 
C_M(Y,Z)\propto\widetilde{c}\int du du' \delta'(u-u')=0.
\eea 
\item The non-local terms in \eqref{nonmm},\ \eqref{nonmq} and \eqref{nonqq} have no contribution.
Since $\dot Y=0$, the function vanishes
\be 
\Delta(\dot Y;\Sigma;u',\Omega)=0.
\ee The non-local terms in \eqref{nonmq} and \eqref{nonmm} are zeros obviously. The non-local term of \eqref{nonqq} is 
\bea 
&&2i\int du du'd\Omega\alpha(u'-u)g_1(\Omega)g_2(\Omega):\Sigma(u,\Omega)\Sigma(u',\Omega):\nn\\&=&2i\int du du'd\Omega \alpha(u-u')g_1(\Omega)g_2(\Omega):\Sigma(u',\Omega)\Sigma(u,\Omega):\nn\\&=&-2i\int du du'd\Omega \alpha(u'-u)g_1(\Omega)g_2(\Omega):\Sigma(u,\Omega)\Sigma(u',\Omega):\nn\\&=&0.
\eea At the second line, we exchanged the variables $u\leftrightarrow u'$. At the third line, we used the antisymmetry of the $\alpha$ function and the symmetry of the normal ordered operator 
\bea 
\alpha(u'-u)=-\alpha(u-u'),\quad :\Sigma(u,\Omega)\Sigma(u',\Omega):\ =\ :\Sigma(u',\Omega)\Sigma(u,\Omega):.
\eea 
\end{itemize}

\section{Light-ray operator formalism}\label{Light-ray}
In this appendix,  we give a review about basic concepts in light-ray operator formalism \cite{Hofman:2008ar,Kravchuk:2018htv} and its relation to our formalism. The concentration is $4$-dimensional conformal field theory. We will first introduce the light-ray operators, and then show that our commutator $[\widetilde{\mathcal{T}}_{\omega,\ell,m},\widetilde{\mathcal{T}}_{\omega',\ell',m'}]$ of the operator $\widetilde{\mathcal{T}}_{f}$ defined in \eqref{wdtt} with $f=e^{-i\omega u}Y_{\ell,m}(\Omega)$ is equivalent to the commutator of $\omega$-deformed energy flow operators in light-ray formalism.

Consider a primary operator ${O}_{\mu_1\mu_2\cdots\mu_s}$ with conformal weight $\Delta$ and spin $s$ which is symmetric and traceless, 
%\be 
%O_{\mu_1\mu_2\cdots\mu_s}=O_{(\mu_1\mu_2\cdots\mu_s)},\quad O^\alpha_{\ \alpha\mu_3\mu_4\cdots\mu_s}=0.
%\ee 
we may introduce a null polarization vector $z^\mu$ and contract it with the $O$ to form an  indices free operator which is a polynomial in $z$
\bea 
O(x,z)=O_{\mu_1\mu_2\cdots\mu_s}(x,z)z^{\mu_1}z^{\mu_2}\cdots z^{\mu_s},\quad z^2=0.
\eea 

This operator $O(x,z)$ is homogeneous in $z$ with degree $s$ 
\be 
O(x,\lambda z)=\lambda^s O(x,z).
\ee %To recover the symmetric traceless tensor $O_{\mu_1\mu_2\cdots\mu_s}$, one may use the following Todorov operator \cite{} 
%\bea 

%\eea 
The light-ray operator ${\bf L}[O]$ is the light transform of operator $O(x,z)$ 
\bea 
{\bf L}[O](x,z)=\int_{-\infty}^\infty d\alpha (-\alpha)^{-\Delta-s}O(x-\frac{z}{\alpha},z)
\eea along the null direction of  $z^\mu$. The light-ray operator transforms as a primary operator with conformal dimension $1-s$ and spin $1-\Delta$.

To implement the light transform, one may introduce the embedding formalism \cite{Dirac:1936,1969AnPhy..53..174M,1970PhRvD...2..293B,Ferrara:1973,1973AnPhy..76..161F,2010JHEP...03..133C,2010PhRvD..82d5031W,2011JHEP...11..154C}. In this formalism, the Minkowski spacetime $\mathbb{R}^{1,3}$ is a projective null cone in $\mathbb{R}^{2,4}$. We will use capital Latin alphabets $ X=(X^{-1},X^0,\cdots,X^4)$ to  denote the coordinates of $\mathbb{R}^{2,4}$. By introducing the light cone coordinates 
\bea 
X^{\pm}=X^{-1}\pm X^4,\quad X^\mu=X^0,X^1,X^2,X^3,
\eea the inner product of $X$ takes the form 
\bea 
X\cdot X=-X^+X^{-}+X^\mu X_{\mu},
\eea where $X^\mu X_\mu=\eta_{\mu\nu}X^{\mu}X^\nu$. A point in Minkowski spacetime $x^\mu$ corresponds to a null vector $ X=(X^+,X^-,X^\mu)$ in light cone coordinates 
\bea 
 X=X^+(1,x^2,x^\mu),\quad x^\mu=\frac{X^\mu}{X^+}.\label{xXcor}
\eea The indices free operator $O(x,z)$ is lifted to the operator $O(X,Z)$ in the embedding formalism 
\bea 
O(X, Z)=(X^+)^{-\Delta}O(x,z)
\eea where $ Z$ is a null vector which is orthogonal to $X$
\be 
 Z^2=X\cdot Z=0.
\ee The null polarization vector $z^\mu$ can be recovered by the relation 
\be z^\mu=Z^\mu-\frac{Z^+}{X^+}X^\mu.
\ee  The primary operator $O(X,Z)$ in the embedding space has the following properties
\bea 
O(X,Z+\beta X)=O(X,Z),\quad O(\rho X,\sigma Z)=\rho^{-\Delta}\sigma^s O(X,Z),\quad \rho,\sigma>0.
\eea Therefore, the light-ray operator can be written as the form 
\bea 
{\bf L}[O](X,Z)=\int_{-\infty}^\infty d\alpha\ O(Z-\alpha X,-X).
\eea In order to be sensitive to the time-dependent structure of the interesting states, we would like to insert a weight function in the light transform as follows
\begin{align}
  {\bf L}_{{\omega/2}} [O](X,Z) \equiv \int_{- \infty}^{\infty} d \alpha \,e^{-i {\omega} \alpha/2} O(Z-\alpha X,-X)\,.
\end{align}
The insertion of \(e^{-i {\omega} \alpha/2}\) will not only improve the convergence of the integral, but also give the detector a non-trivial null momentum. This transformation is called \(\omega\)-deformed light transform in \cite{Korchemsky:2021okt}. The resulting operator corresponds to generalized event shapes.

To approach null infinity along the direction of a null vector \(n^\mu=(1,n^i)\) in Cartesian coordinates \cite{Gonzo:2020xza}, we consider the following series of points\footnote{In embedding formalism, we use light cone coordinates to express the vectors of $\mathbb{R}^{2,4}$ from now on.}
\begin{align}
  X_{v}= \left(0,1 , {\bar n^\mu \over 2 v}\right)\,, \qquad
  Z_{v} = \left({2 \over v}, 0 , n^\mu\right),
\end{align} 
with \(\bar n=(-1,n^i)\) a null vector satisfying \(n\bar n=2\), and \(v=t+r\) the advanced time. It is easy to see that $Z_v^2 = X_v^2 =Z_v \cdot X_v = 0$. It follows that
\begin{align}
  Z_v-\alpha X_v=\left(\frac{2}{v},{-\alpha},n^\mu-{\alpha\bar n^\mu \over  2 v}\right).
\end{align} 
Compared with \eqref{xXcor}, the null  vector $Z_v-\alpha X_v$ corresponds to the point 
\bea 
x^\mu=\frac{v}{2}n^\mu-\frac{u}{2}\bar{n}^\mu=(t,rn^i),
\eea when the parameter $\alpha$ is related to the retarded time as follows
\be \alpha=2u.
\ee 
Taking the limit of \(v\to\infty\) while keeping $u$ finite, and considering the conformal property of \(O(X,Z)\), we get
% In Minkowski spacetime, a null ray at null infinity can be represented by a unit vector \(n^\mu=(1,\vec n)\), with \(\vec n\) a unit normal vector on \(S^{2}\). In our retarded spherical coordinates, we can rewrite it as \(n^\mu=(0,1,\theta^A)\). The so-called light-ray operators are of form 
\begin{align}
  \label{calO-def}
  \mathcal O_{\omega,s}(n) \equiv& 2^{s-1}\lim_{v \to \infty} {\bf L}_{{\omega}/2} [O](X_v,Z_v) \nn \\
  =&2^{s-1}\lim_{v \to \infty}\int_{-\infty}^\infty d\alpha e^{ -i \alpha \omega /2}O(Z_v-\alpha X_v,-X_v)\nn\\
  =&2^{-1}(-1)^s\lim_{v \to \infty}\int_{-\infty}^\infty d\alpha e^{ -i \alpha \omega /2}\left(\frac{2}{v}\right)^{-\Delta}O_{\mu_1\cdots\mu_s}\left(x^\mu\right){\bar n^{\mu_1} \over v  }\cdots{\bar n^{\mu_s} \over v  }\nn\\
  =&2^{-\Delta}(-1)^s\lim_{v \to \infty}v^{\Delta-s}\int_{-\infty}^\infty du e^{ -i \omega u}O_{\mu_1\cdots\mu_s}\left(x^\mu\right)\bar n^{\mu_1} \cdots  \bar n^{\mu_s},
\end{align}

\iffalse and then obtain
\begin{align}
  \mathcal O_{\omega,s}(n){=}& 2^{s-\Delta}(n\bar n)^{1-s} \lim_{v\to\infty} v^{{\Delta} - s} \int_{-\infty}^\infty d\alpha e^{ -i \alpha \omega (n\bar n)}\, O_{\mu_1\cdots\mu_s} \left(\frac{vn^\mu}{2}+ \alpha \bar n^\mu\right)\bar n^{\mu_1} \cdots  \bar n^{\mu_s} ,
\end{align}
after rescaling \(\frac{\alpha}{2(n \bar n)}\to\alpha\). Now, taking \(\alpha=-u/2\) and \(\bar n^\mu=(-1,n^i)\), we find
\begin{align}
  \mathcal O_{\omega,s}(n){=}& -2^{-\Delta}\lim_{v\to\infty} v^{{\Delta} - s} \int_{-\infty}^\infty du e^{ i \omega u}\, O_{\mu_1\cdots\mu_s} \left(t,x^i\right)\bar n^{\mu_1} \cdots  \bar n^{\mu_s}.
\end{align}\fi

 In light cone coordinates $(u,v,\theta^A)$, the above null vector becomes
\be 
\bar n^\mu=(-2,0,0,0).
\ee So the light-ray operator $\mathcal O_{\omega,s}(n)$ takes form
\bea 
\mathcal O_{\omega,s}(n)= \lim_{r\to\infty} r^{{\Delta} - s} \int_{-\infty}^\infty du e^{- i \omega u}\, O_{\underbrace{u\cdots u}_{s}}(x).
\eea 
The energy flow operator \(\mathcal{E}_\omega (n)\) is the light-ray operator of stress-energy tensor \(\widetilde{T}_{\mu_1\mu_2}\) with conformal weight \(\Delta=4\) and spin \(s=2\). Namely, 
\begin{align}\label{calE-def}
  \mathcal E_{\omega}(n) =\lim_{r\to\infty}r^2 \int_{-\infty}^\infty du e^{- i \omega u}\, \widetilde{T}_{uu}.
\end{align}
In particular, the soft limit \(\omega=0\) gives the famous average null energy operator. In free scalar theory, the symmetric traceless stress-energy tensor is 
\begin{align}
  \widetilde{T}_{\mu\nu}=\partial_\mu\Phi\partial_\nu\Phi-\frac{1}{6}\partial_\mu\partial_\nu\Phi^2-\frac{1}{12}\eta_{\mu\nu}\partial^2\Phi^2,
\end{align}

% Similar to the fall-off condition in the case of retarded coordinates, we require
% \begin{align}
%   \Phi(t,x^i)=\frac{\Sigma(u,\Omega)}{v}+\co(v^{-2})
% \end{align}
Considering the fall-off condition of scalar field \(\Phi\), we obtain
\begin{align}
  \mathcal E_{\omega}(\Omega)=\int_{-\infty}^\infty du e^{- i \omega u}\left[:\dot\Sigma^2(u,\Omega):+\frac{1}{6}\omega^2:\Sigma^2(u,\Omega):\right].
\end{align} 
It is easy to see the equivalence between this operator and the aforementioned $\widetilde{\mathcal{T}}_f$. To be more accurate, we could insert a spherical harmonic function $Y_{\ell,m}(\Omega)$ as the weight function about angular coordinates, and then impose integration on energy flow operator with respect to $\Omega$. These operations lead to the following identification
\begin{align}
    \widetilde{\mathcal{T}}_{\omega,\ell,m}\equiv\int d\Omega Y_{\ell,m}(\Omega)\mathcal{E}_{\omega}(\Omega)=\int_{-\infty}^\infty dud\Omega e^{- i \omega u}Y_{\ell,m}(\Omega)\left[:\dot\Sigma^2(u,\Omega):+\frac{1}{6}\omega^2:\Sigma^2(u,\Omega):\right].
\end{align} Using the definition of the smeared operators in the context and going back to the position space, this is exactly
\bea 
\widetilde{\mathcal{T}}_{f}=\mathcal T_{f}-\frac{1}{6}\mathcal{Q}_{\ddot f},
\eea where the function $f=f(u,\Omega)$. Interestingly, the light-ray transform of the scalar operator $\Phi^2$ with $\Delta=2$ and $s=0$ is 
\bea 
\int_{-\infty}^\infty du e^{-i\omega u}:\Sigma^2(u,\Omega):.
\eea This corresponds to the smeared operator $\mathcal{Q}_g$ in the context. One can further check that the commutator of $\widetilde{\mathcal{T}}_{\omega,\ell, m}$ is
\begin{align}
  &[\widetilde{\mathcal{T}}_{\omega,\ell, m},\widetilde{\mathcal{T}}_{\omega',\ell', m'}]=(\omega'-\omega)\sum_{L=|\ell-\ell'|}^{\ell+\ell'}\sum_{M=-L}^{L}c_{\ell, m; \ell', m'; L, M}\widetilde{\mathcal{T}}_{\omega+\omega', L, M}- (-1)^m\frac{\omega^3}{12}c\delta(\omega+\omega')\delta_{\ell,\ell'}\delta_{m,-m'}\nn\\
 % &+\frac 1{18}c\int du du'd\Omega[\beta(u-u')^2-\beta(u'-u)^2]\omega^2\omega'^2e^{-i(\omega u+\omega'u')}\label{tildett1}\\
  &+\frac{i}{36}\int du du'd\Omega \alpha(u-u')\omega^2\omega'^2e^{-i(\omega u+\omega'u')}Y_{\ell,m}(\Omega)Y_{\ell',m'}(\Omega)[\Sigma(u, \Omega)\Sigma(u', \Omega)+\Sigma(u',\Omega)\Sigma(u,\Omega)].\nn
\end{align}
It matches with the corresponding commutator in \cite{Korchemsky:2021htm}.

\bibliography{refs}

\end{document}